\documentclass[aps,prx,onecolumn]{revtex4-2} 
\usepackage{graphicx}
\usepackage{amsmath,amsfonts,amssymb}
\usepackage{bm}
\usepackage{color}

\usepackage{xcolor}

\usepackage[normalem]{ulem}

\usepackage{soul}

\begin{document}

\title{Fluctuation-Dissipation Relations in Active Matter Systems}

\author{Lorenzo Caprini} 
\affiliation{School of Sciences and Technology, University of Camerino, Via Madonna delle Carceri, I-62032, Camerino, Italy.}
\author{Andrea Puglisi} 
\affiliation{Institute for Complex Systems - CNR, P.le Aldo Moro 2, 00185, Rome, Italy.}
\affiliation{Department of Physics, University of Rome Sapienza, P.le Aldo Moro 2, 00185, Rome, Italy.}
\affiliation{ INFN, University of Rome Tor Vergata, Via della Ricerca Scientiica 1, 00133, Rome, Italy.}
\author{Alessandro Sarracino}
\affiliation{Department of Engineering, University of Campania ``Luigi Vanvitelli'',  81031, Aversa (Caserta), Italy.}

\date{\today}

\begin{abstract}
We investigate the non-equilibrium character of self-propelled particles through the study of the linear response of the active Ornstein-Uhlenbeck particle (AOUP) model. We express the linear response in terms of correlations computed in the absence of perturbations, proposing a particularly compact and readable fluctuation-dissipation relation (FDR): such an expression explicitly separates equilibrium and non-equilibrium contributions due to self-propulsion. As a case study, we consider non-interacting AOUP confined in single-well and double-well potentials. In the former case, we also unveil the effect of dimensionality, studying one, two, and three-dimensional dynamics. We show that information about the distance from equilibrium can be deduced from the FDR, putting in evidence the roles of position and velocity variables in the non-equilibrium relaxation.
\end{abstract}

\maketitle

\section{Introduction}

The fluctuation-dissipation relations (FDR) played a pivotal role in the development of non-equilibrium statistical mechanics. Starting from the regime of weak departure from equilibrium, with the pioneering work of Einstein on mobility-diffusivity relation, through the central work of Onsager on the reciprocal relations~\cite{O31}, a unification of further theoretical results was obtained in the Kubo's linear response theory~\cite{K57}, which includes also Green-Kubo relations useful for transport coefficients in spatially extended systems. In Onsager's and Kubo's theories, all relations remain ``simple'', provided that some kind of time-reversal symmetry holds in the unperturbed state. As soon as this symmetry is lost, the simplicity of linear response and transport coefficients is no more guaranteed~\cite{marconi2008fluctuation}.

Time-reversal symmetry is broken in widespread natural phenomena: a
constraint against equilibration is usually given by non-equilibrium
boundary conditions, or in the presence of very slow relaxation
dynamics, as in glassy systems.  Hindrance to equilibration can also
be caused by internal non-conservative forces, such as those in
granular systems and in self-propelled particles. In all those cases
response theory must be generalized, paying a price in simplicity.
Several generalised relations have been proposed in the last decades:
they still connect the perturbed system to the unperturbed one, but
often require some detailed microscopic knowledge of the system. Some
generalised relations may take a simple form in particular situations,
e.g. an effective temperature may replace the thermostat temperature
in a range of well-separated timescales for aging glassy systems, but
this scenario is far from being
general~\cite{cugliandolo2011effective,puglisi2017temperature}. More
frequently, one faces a situation where the equilibrium FDR is
modified by additional contributions of a more complex nature.

The several approaches to FDR present in the literature belong to two main classes. Class A requires the knowledge of the stationary distribution, while class B only requires the knowledge of the microscopic dynamical rules (transition rates or Langevin equations). First examples of FDR of class A were obtained for chaotic systems and Brownian dynamics with non-conservative forces~\cite{A72,FIV90}. In those cases, the knowledge of the steady-state probability distribution (at least, in some approximations) still maintains a fundamental importance: in the absence of this information the generalized FDR remains, somehow, an implicit relation which can be useful to test ansatz on the steady-state properties~\cite{gnoli2014}. More recent FDR of this kind express the non-equilibrium contributions in terms of stochastic entropy production~\cite{ss06, seifert2010fluctuation}.
Examples of FDR of class B are derived from the {\em Malliavin weight sampling}~\cite{malliavin}, or the Novikov theorem~\cite{n65}, which have been intensively employed in the context of glassy physics and represent a powerful tool to calculate the susceptibility and, thus, the so-called effective temperature of a system~\cite{cugliandolo2011effective,CKP94}. However, in this approach, the FDR involves correlations with the noise and the physical meaning of these terms remain often difficult to catch.
Other schemes in this class include the description in terms of ``frenetic'' contributions~\cite{baiesi2009fluctuations, maes}, focussing on the role of time-symmetric fluctuations out of equilibrium (see also~\cite{PhysRevE.78.041120} for derivations of analogous FDR for discrete spin variables).

The generalization of the FDR to active matter represents a challenging issue. Systems of active particles are usually far from equilibrium and many of them move in the solvent through complex mechanisms involving chemical reactions or mechanical agents, such as cilia or flagella~\cite{marchetti2013hydrodynamics, bechinger2016active, elgeti2015physics, gompper20202020}. In the spirit of minimal modelling, these systems could be described through simple stochastic dynamics that resembles that of passive colloids except for the addition of a coarse-grained time-dependent force, often called self-propulsion or simply active force~\cite{shaebani2020computational,fodor2018statistical}. 
This force replaces the microscopic details of the system that, in general, are related to complicate internal mechanisms of energy transduction and involve an intrinsic source of strong deviation from thermodynamic equilibrium. Except for special cases, the steady-state properties of these non-equilibrium models are not known analytically or without approximations and, thus, FDR of class A remain in implicit forms~\cite{caprini2018linear, sarracino2019fluctuation}. Explicit expressions can be worked out in the limit of small persistence time $\tau$, the first example obtained in~\cite{fodor2016far}. On the other hand, the Malliavin weight sampling procedure has been applied to the case of active particle dynamics~\cite{szamel2017evaluating} and, in particular, employed to numerically calculate susceptibility and effective temperature of suspensions of active particles~\cite{berthier2013non, levis2015single, nandi2018effective, cugliandolo2019effective, preisler2016configurational}, even, in phase-separated configurations~\cite{petrelli2020effective}. We remark that usually the concept of effective temperature is thermodynamically meaningful only in systems with well-separated time-scales~\cite{VBPV09}. 
This approach has been also employed to calculate the transport coefficients, such as the mobility, in combination with an approximate method valid at low-density values or small persistence regimes~\cite{dal2019linear, cengio2020fluctuation}.
An attempt to generalise FDR to active systems has been recently presented in~\cite{brady} and in~\cite{maes2020fluctuating, maes}: in the latter case a relation involving a double-time derivative of correlators appears, making the meaning of the formula and its numerical (and experimental) application not immediate.


In this paper, we provide a clear example in the framework of active matter for which a generalized FDR of class B can be explicitly obtained in terms of simple steady-state correlations, involving functions of the observables (such as positions and velocities), which do not require a knowledge of the steady-state properties and do not involve any approximation (e.g. for small persistence regimes, etc.). Moreover, we propose a new quantity to measure the departure from equilibrium, which is defined in terms of the generalized FDR and takes into account the features of the system under time-reversal. We study the behavior of such a quantity for non-interacting self-propelled particles confined via different potentials, ranging from non-harmonic single-well to double-well potentials. In the former case, we also investigate the effect of the dimensionality on the non-equilibrium features.

The article is structured as follows: in Sec.~\ref{sec:model}, we introduce model and notations, while in Sec.~\ref{sec:FDR} we report the FDR relation obtained in our approach.
Sec.~\ref{sec:numerics} is dedicated to the numerical study of the linear response, exploring both convex and non-convex potentials, in one, two, and three dimensional systems. Finally, in Sec.~\ref{sec:concl}, we compare our results with two popular approximated results showing their failure, while the final section is dedicated to discussions and conclusions.

\section{Self-propelled particles}
\label{sec:model}
We consider a well-known scheme to describe the behavior of self-propelled particles, the Active Ornstein-Uhlenbeck (AOUP)
model~\cite{berthier2017active, mandal2017entropy, caprini2018active, wittmann2018effective, bonilla2019active, dabelow2019irreversibility, martin2020statistical, woillez2020active} (also known as Gaussian Colored Noise (GCN)).
The AOUP has been employed to reproduce the phenomenology of passive
colloids immersed in a bath of active particles~\cite{wu2000particle,
  maggi2014generalized, maggi2017memory} but also - perhaps at a more approximate level - the dynamics of self-propelled
particles themselves. 
Its connection with other popular
models for self-propelled particles has been addressed by some authors~\cite{caprini2019comparative, das2018confined}.  For instance, the AOUP model
can reproduce the accumulation near the boundaries of channels and obstacles~\cite{caprini2018active, caprini2019active}, the non-equilibrium clustering or phase-separation~\cite{fodor2016far, farage2015effective, maggi2020universality} typical of active matter and
the spatial velocity correlation spontaneously observed in dense
active systems~\cite{caprini2020hidden, caprini2020time}.
According to the AOUP scheme, the position in $d$ dimensions
$\mathbf{x}$ of the self-propelled particle evolves with the
following stochastic equation:
\begin{equation}\label{eq:motion}
\gamma\dot{\mathbf{x}}(t) =\mathbf{F} (t)+ \mathbf{f}_a(t) + \mathbf{h}(t) \,,
\end{equation}
where $\gamma$ is the drag coefficient. In Eq.~\eqref{eq:motion}, we
have neglected the thermal noise due to the solvent since the thermal
diffusion is usually smaller than the effective diffusion due to the
active force, in several experimental active systems~\cite{bechinger2016active}.  The term $\mathbf{F}(t)= - \nabla
U[\mathbf{x}(t)]$ models the force due to an external confining
potential, while the term $\mathbf{f}_a(t)$ represents the
self-propulsion of the particle $i$ and is described by an
Ornstein-Uhlenbeck process: 
\begin{equation}\label{eq:motion2}
\tau\dot{\mathbf{f}}_a = -\mathbf{ f}_a + \gamma \sqrt{2 D_a} {\boldsymbol{\xi}} \,,
\end{equation}
where ${\boldsymbol{\xi}}$ is a white noise vector with zero average
and unit variance. The parameter $\tau$ is the persistence time while $D_a$ is the diffusion coefficient due to the self-propulsion.
Finally, the term $\mathbf{h}(t)$ is an external small perturbation used to probe the linear response properties of the system. In what follows, the perturbed dynamics will be denoted by the superscript $h$ to be distinguished by the unperturbed one. The response function can be computed as the observed (normalised) variation of some observable when the perturbation is an impulse at time $0$, i.e. $\mathbf{h}(t)=\gamma \delta \mathbf{x}(0) \delta(t)$ (with $\delta\mathbf{x}(0)$ a vector with infinitesimal displacements representing the impulsive perturbation). 
In general, with the exception of singular measures, it is possible to derive a compact formula for the response function of one of the degrees of freedom $x_i$ of the system to the perturbation of a degree of freedom $x_j$, $\mathcal{R}_{x_ix_j}$, defined - in the steady-state - as
\begin{equation}
\label{eq:numericalresponse}
\mathcal{R}_{x_ix_j}(t) = \frac{\langle x_i(t)\rangle^h-\langle x_i(t)\rangle}{\delta x_j(0)}
=\gamma\left.\frac{\delta\langle x_i(t)\rangle^h}{\delta h_j(0)}\right|_{h=0} \,,
\end{equation}
where $t>0$ and, in the last equality, we have introduced the functional derivative with respect to the perturbative force $h_j(0)$.
In the following, for the one-dimensional case, we will replace $\mathcal{R}_{xx}(t)$ with $\mathcal{R}_{x}(t)$ for notational convenience.

\section{Generalized fluctuation-dissipation relation for self-propelled particles}
\label{sec:FDR}

The model introduced so far reaches a non-equilibrium stationary state even in the absence of perturbation displaying a non-vanishing entropy production~\cite{fodor2016far, puglisi2017clausius, caprini2019entropy, dabelow2020irreversibility}, with a few special exceptions: in the absence of external potential, or in the presence of linear forces, the detailed balance is restored and the system recovers an apparent equilibrium state with vanishing entropy production~\cite{marconi2017heat}. This is an imperfection of the AOUP model \cite{caprini2018comment}. Except for these special cases, the probability distribution, $p(\mathbf{x}, \mathbf{f}_a)$, is unknown. An approximation can be obtained, valid in the proximity of equilibrium ($\tau=0$)~\cite{fodor2016far, marconi2017heat, bonilla2019active, martin2020aoup}. Therefore, in general, this model of self-propelled particles cannot be treated within the FDR of class A, because the FDR remains implicit:
\begin{equation}
\label{eq:response_vulpio}
\mathcal{R}_{x_ix_j}(t)= - \left\langle x_i(t) \frac{\partial}{\partial x_j} \log [p(\mathbf{x}(0), \mathbf{f}_a(0))] \right\rangle \,,
\end{equation}
where the average $\langle \cdot\rangle$ is obtained considering the unperturbed system and $p(\mathbf{x}, \mathbf{f}_a)$ is the stationary probability distribution in the absence of perturbations that depends on the whole set of particle positions and velocities.
Eq.~\eqref{eq:response_vulpio} can be used only perturbatively in powers of the persistence time since the expression of $p(\mathbf{x}, \mathbf{f}_a)$ is only known perturbatively in powers of $\tau$ \cite{caprini2018linear}. For small $\tau$, the system is near the equilibrium, and, thus, the relation~\eqref{eq:response_vulpio} is restricted to the same regimes. A similar approach has been employed in~\cite{fodor2016far}.

We propose an alternative approach, inspired by FDR of class B, to
calculate the response function in terms of suitable steady-state
correlations.
Below, we report the final outcome of our technique applied to an AOUP
confined by an external potential, while the detailed calculations are
accurately described in Appendix~\ref{App:Novikov}:
\begin{flalign}
\label{eq:second_result}
D_a\gamma\mathcal{R}_{x_i x_j}(t) =&
\frac{1}{2}\left[ \left\langle x_i(t) \nabla_{j} U(0) \right\rangle +\left\langle \nabla_{i} U(t) x_j(0) \right\rangle  \right]\\
&+\frac{\tau^2}{2}\left[ \left\langle v_k(t) \nabla_{k}\nabla_{i} U(t) v_j(0) \right\rangle +\left\langle v_i(t) \nabla_{j}\nabla_{k} U(0) v_k(0) \right\rangle  \right] \,,
\end{flalign}
where all the correlations have been evaluated in the steady-state and we have introduced the particle velocity $v_j=\dot{x}_j$.
According to our notation, repeated indices are summed, $U(s)=U(\mathbf{x}(s))$ and $\nabla_i = \partial_{x_i}$.
Let us comment in detail on Eq.~\eqref{eq:second_result}: first, we observe that it is symmetric under time reversal (i.e. when times $t$ and $0$ are swapped) and involves both position and velocity correlations.
The first line of Eq.~\eqref{eq:second_result} has the same form of the equilibrium FDR holding for passive particles.
Indeed, when the detailed balance condition holds (for $\tau \to 0$), the first and the second terms on the right-hand side of Eq.~\eqref{eq:second_result} coincide, while the second line disappears,
in such a way that $D_a\gamma\mathcal{R}_{x_i x_j}=\left\langle
x_i(t) \nabla_j U(0) \right\rangle$. The second line provides two additional terms that are exquisitely non-equilibrium contributions to the response function. Remarkably, these non-equilibrium correlations involve the particle velocity, without containing direct simple dependence on the perturbed observable, namely the particle position (but of course position is involved through the potential). At variance with the well-known equilibrium scenario, in active systems, $\mathcal{R}_{x_i x_j}(t)$ is not only determined by a time correlation involving the position but is strongly
affected by the correlations between the other variables involved in
the dynamics, the velocity in this case. The harmonic case - which is
a special case where the AOUP model satisfies the detailed balance even when $\tau>0$ - is treated in Appendix~\ref{App:Harmonioscillator}.

Understanding what are the leading terms in Eq.~\eqref{eq:second_result} is a challenging issue that cannot be performed in general but requires a numerical analysis. In addition, since the detailed balance does not hold in systems of active particles, in general, the reversibility is broken by the presence of non-zero steady-state currents that produce non-vanishing entropy. The structure of Eq.~\eqref{eq:second_result} suggests a way to measure how much the system is far from equilibrium by quantifying the unbalance between the couples of terms, to see the effect of irreversibility on the response function. Therefore, we define:
\begin{flalign}
\label{eq:xU_diff}
&D_a\gamma\mathcal{D}^x_{ij}(t)=\left\langle x_{i}(t) \nabla_j U(0) \right\rangle - \left\langle \nabla_i U(t) x_j(0) \right\rangle\\
\label{eq:vU_diff}
&D_a\gamma\mathcal{D}^v_{ij}(t)=\tau^2\left\langle v_k(t) \nabla_{ k} \nabla_{ i} U(t) v_j(0) \right\rangle - \tau^2\left\langle v_i(t) \nabla_{j} \nabla_{k}  U(0) v_k(0) \right\rangle \,,
\end{flalign}
where all the correlations are calculated in the steady-state. $D^x_{ij}(t)$ and $D^v_{ij}(t)$ vanish at the initial time, $t=0$ and for any $t$ in any equilibrium configurations where the detailed balance holds.
For notational convenience, we replace $D^x_{ij}$ and $D^v_{ij}$ with $D_x$ and $D_v$ in the one-dimensional case.

\section{Numerical results}
\label{sec:numerics}

In this section, we numerically check the FDR~\eqref{eq:second_result}, in two typical cases that cannot be
analytically solved: A) the quartic potential, B) the double-well
potential.  On the one hand, A) represents the simpler convex case of
study, except for the harmonic confinement where correlation functions
and responses can be analytically computed.  In the harmonic case, we
remark that the expression~\eqref{eq:second_result} is consistent with
previous results~\cite{szamel2014self, caprini2018linear} that, in
particular, predict an exponential time-decay of the response function
with a typical time $\gamma/k$ (see the Appendix~\ref{App:Harmonioscillator}).  In this simple case, the response
function does not show any dependence on the parameters of the active
force.  On the other hand, B) is the simpler non-convex case able to
reveal the interplay between the self-propulsion and the non-convex
curvature of the potential. This feature has already manifested many
dynamical anomalies leading to the emergence of regions with effective
negative mobility in the large persistence regime~\cite{caprini2019activedoublewell} showing a bifurcation-like behavior
with non-local stationary probability distribution~\cite{woillez2020nonlocal}.  In the quartic case, we also unveil the
role of the dimensions comparing the results for one, two and three
dimensional systems.

The numerical study is performed keeping fixed $D_a$ and the potential strength, to investigate the role of the persistence time, $\tau$. Time is measured in units of a typical time $t^*$, ruling the relaxation of the response function for the passive system (obtained for $\tau \to 0$), which is chosen as a reference case. In particular, $t^*=\gamma/k$ for the harmonic potential and $t^*=1/\sqrt{D_a k}$ for the quartic case. In all the cases, the system is perturbed via a small force along the $x$-axis, with amplitude $\delta x=10^{-3}$. This choice guarantees the linear regime of the response calculated using Eq.~\eqref{eq:numericalresponse}.

\subsection{Quartic potential}

\subsubsection{One-dimensional system}

\begin{figure}[!t]
\centering
\includegraphics[width=0.99\linewidth,keepaspectratio]{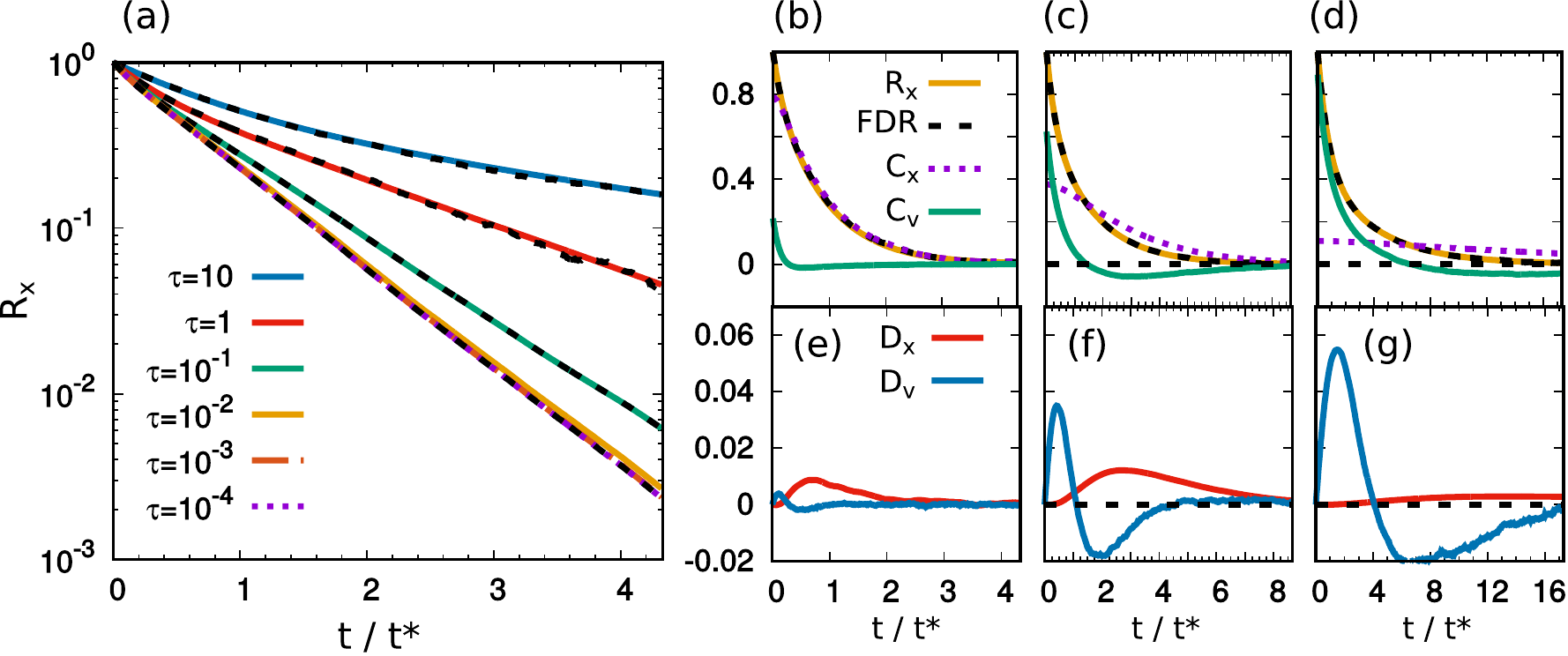}
\caption{\label{fig:responsevstime1} Panel (a): Response function,
  $R_{x}(t)$, for different values of $\tau$ calculated numerically
  via Eq.~\eqref{eq:numericalresponse} in the case of a quartic
  potential, $U(x)=k x^4/4$.  Times are measured in units of the typical
  time $t^*$ (see main text).  The dashed
  black lines are obtained by using the FDR,
  Eq.~\eqref{eq:second_result}.  Panels (b), (c), (d): $R_x(t)$ (solid
  yellow lines) compared to the FDR (dashed black lines) ,
  Eq.\eqref{eq:second_result}.  Dotted violet lines represent
  $C_x=\langle x(t) U'(0)\rangle/(D_a\gamma) + \langle U'(t)
  x(0)\rangle/(D_a\gamma)$ while green solid lines represent
  $C_v=\tau^2\langle v(t) U''(0) v(0)\rangle/(D_a\gamma) + \tau^2\langle
  v(t) U''(t) v(0)\rangle/(D_a\gamma)$.  Panels (e), (f), (g): $D_x=\langle
  x(t) U'(0)\rangle/(D_a\gamma) - \langle U'(t) x(0)\rangle/(D_a\gamma)$ (solid
  red lines) and $D_v=\tau^2\langle v(t) U''(t) v(0)\rangle/(D_a\gamma)
  -\tau^2\langle v(t) U''(0) v(0)\rangle/(D_a\gamma)$ (solid blue lines).
  Panels (b),(e) are obtained with $\tau=10^{-1}$, panels (c), (f)
  with $\tau=1$ and, finally, panels (d), (g) with $\tau=10$.  The
  other parameters are $D_a=1$, $k=3$, $\gamma=1$ }
\end{figure}

Panel (a) of Fig.~\ref{fig:responsevstime1} shows the response function, $\mathcal{R}_{x}$, in
the case of a quartic potential, $U(x)= k x^4 /4$ for several values
of $\tau$ exploring both the small and the large persistence regime,
i.e. the near and far equilibrium regimes, respectively.  The figure
reveals the agreement between $\mathcal{R}_x(t)$ calculated from its definition,
given by Eq.~\eqref{eq:numericalresponse}, and using the correlation
functions appearing in Eq.~\eqref{eq:second_result} adapted to the
one-dimensional case.  This numerically confirms the extension of the
FDR to AOUP active particles, even far from equilibrium.
Specifically, in the small $\tau$ regime, the different curves of
$\mathcal{R}_x(t)$ collapse onto the same curve that corresponds to the response
function of a passive Brownian particle at temperature $D_a\gamma$.
Instead, in the large $\tau$ regime, the larger is $\tau$ the slower
is the relaxation of the response function, $\mathcal{R}_x(t)$, that roughly
displays two decay regimes.  Indeed, it is known that AOUP in a
single-well non-harmonic potential accumulate far from the potential
minimum roughly near the points where the active force is balanced by
the external one, in such a way that the spatial density shows two
symmetric peaks~\cite{caprini2019activity}.  Somehow, the system
behaves as if an effective double-well confinement control the
particle dynamics, an effect captured by many approximated results~\cite{caprini2019activity}.
In the quartic case, the $\tau$-dependence is a non-equilibrium consequence caused by the interplay between the non-linearity of the potential and the persistence  of the active force.
Indeed, in the harmonic case, the response is $\tau$-independent and uniquely determined by the relaxation time of the potential~\cite{szamel2014self}, namely $t^*=k/\gamma$.

%
\begin{figure}[!t]
\centering
\includegraphics[width=0.99\linewidth,keepaspectratio]{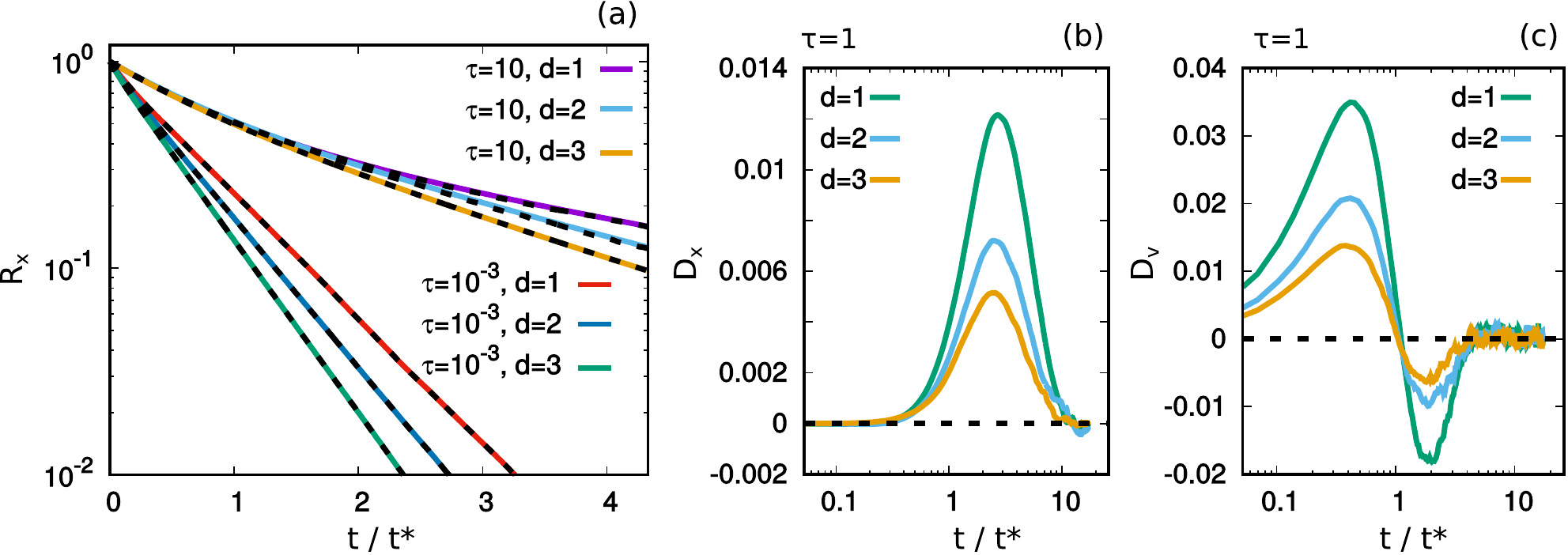}
\caption{\label{fig:moredimensions} Panel (a): Response function,
  $R_{x}(t)$, for two different values of $\tau=10^{-3}, 10$ and
  dimensions $d=1,2,3$, calculated numerically via
  Eq.~\eqref{eq:numericalresponse}, in the cases of quartic potentials
  $U(\mathbf{x})=k |\mathbf{x}|^4/4$.  The dashed black lines are
  obtained by using the FDR, Eq.~\eqref{eq:second_result}.  Panel (b)
  and (c): $D_x=\langle x(t) U'(0)\rangle/(D_a\gamma) - \langle U'(t)
  x(0)\rangle/(D_a\gamma)$ (panel (b)) and $D_v=\tau^2\langle v(t) U''(t)
  v(0)\rangle/(D_a\gamma) - \tau^2\langle v(t) U''(0) v(0)\rangle/(D_a\gamma)$
  (panel (c)) for $d=1,2,3$, at $\tau=1$.  The other parameters are
  $D_a=1$, $k=3$, $\gamma=1$. }
\end{figure}

In Fig.~\ref{fig:responsevstime1}, panels~(b)-(g), the terms of the FDR
(Eq.~\eqref{eq:second_result}) are separately studied and compared with
$\mathcal{R}_{x}(t)$, for different values of $\tau=10^{-1}, 1, 10$,
respectively.  As emerges from Fig.~\ref{fig:responsevstime1} (b)-(d),
the equilibrium terms, $D_a \gamma C_x(t)=\langle x(t) U'(0)\rangle + \langle
U'(t) x(0)\rangle$, are dominant in the near-equilibrium
configurations, for small values of $\tau$. In these cases, the
non-equilibrium velocity correlations, i.e. $D_a \gamma C_v(t)=\tau^2\langle v(t) U''(0)
v(0)\rangle + \tau^2\langle v(t) U''(t) v(0)\rangle$,
only weakly affect the first time decay of $\mathcal{R}_x(t)$ also vanishing
approximatively at $t \sim \tau$.  When $\tau$ is increased, the
contribution of the velocity correlation starts growing even if its
decay remains roughly determined by $\tau$, while the potential
correlation decreases slowly.  Interestingly, the velocity correlation
reaches negative values and increases very slowly towards zero almost
balancing the potential correlation, as shown in
Fig.~\ref{fig:responsevstime1}~(b)-(d).  The term $D_a \gamma C_v(t)=\tau^2\langle v(t) U''(0)
v(0)\rangle + \tau^2\langle v(t) U''(t) v(0)\rangle$,
becomes dominant in the large persistence regime, as shown in
Fig.~\ref{fig:responsevstime1}~(b)-(d), while the term $D_a \gamma C_x(t)=\langle x(t)
U'(0)\rangle + \langle U'(t) x(0)\rangle$ gives a
negligible contribution.

Finally, to understand the role of the detailed balance violation in
the temporal decay of the response, we plot the terms
$\mathcal{D}_x(t)$ (solid red lines) and $\mathcal{D}_v(t)$ (solid
blue lines), in panels Fig.~\ref{fig:responsevstime1}~(e)-(g). If these
two terms are close to zero, then the detailed balance holds.  As
expected, this occurs for small $\tau$ while the detailed balance is
strongly violated as far as $\tau$ is increased.  Moreover, the
irreversibility manifests in different ways: $\mathcal{D}_x(t)$ is
positive and reaches a peak for $t \sim \tau$, while
$\mathcal{D}_v(t)$ shows an oscillation from positive to negative
values.
As a final remark, the violation of the detailed balance strongly
suggests that any equilibrium-like approaches that assume this
condition, such as the Unified Colored Noise approximation (that will be explicitly evaluated in Sec.~\ref{Sec:V}), cannot work in the large persistence regime.

\subsubsection{Higher dimensional systems}

We explore the role of the system dimensions, both on response
functions and FDR.  In the case of harmonic confinement (in
equilibrium both for dimensions $d=1,2,3$), where the detailed balance
holds, the dimensions of the system do not affect the time-decay of
the response function.

In Fig.~\ref{fig:moredimensions}~(a), we compare $\mathcal{R}_x(t)$ 
for one-, two- and three-dimensional self-propelled particles confined
through the quartic potentials, $U(\mathbf{x})=k|\mathbf{x}|^4/4$. We
explore both the small and the large persistence regime reporting two
reference cases, $\tau=10^{-3}, 10$, respectively.  In the small
persistence regime ($\tau=10^{-3}$) that roughly coincides with the
passive case, the larger is the dimension the faster is the decay of
$\mathcal{R}_{x}(t)$, an effect that is entirely due to the non-linearity of the
potential.  The increase of the system dimensions increases the
effective trapping of the particle because of the confinement, an
effect not related to the active force. Instead, in the large
persistence regime ($\tau=10$), the correlations show a first temporal
decay which does not depend on the dimension of the system (that
coincides for $d=1,2,3$), while, in the second stage of the decay, the
higher dimensional systems decrease faster than their corresponding
in lower dimensions. Somehow, the active force can suppress
the dimensional dependence of the response function, at least for a
small time-window.

Fig.~\ref{fig:moredimensions}~(b) and~(c) show the functions
$\mathcal{D}_x(t)$ and $\mathcal{D}_v(t)$ for $d=1,2,3$ at $\tau=1$, for
simplicity. We highlight that results are similar for other values of
$\tau$ (not shown).  The qualitative functional forms of $D_x(t)$ and
$D_v(t)$ are unchanged with $d$, resembling the shape described in the
one-dimensional system: a single peak around $t \sim \tau$ for
$\mathcal{D}_x(t)$ and an oscillation from positive to negative values
for $\mathcal{D}_v(t)$.  The higher is the dimension of the system, the
smaller is the maximal amplitude of both $\mathcal{D}_x(t)$ and $\mathcal{D}_v(t)$.
This means that the increase of $d$, somehow, reduces the breaking of
the detailed balance, producing smaller currents.  This is also
consistent with the increasing difficulties to observe collective
phenomena when the dimensions are increased, that in three dimensions
typically require larger values of the active forces compared to
two-dimensional systems~\cite{stenhammar2014phase}.

\subsection{Double-well potential}

\begin{figure}[!t]
\centering
\includegraphics[width=0.99\linewidth,keepaspectratio]{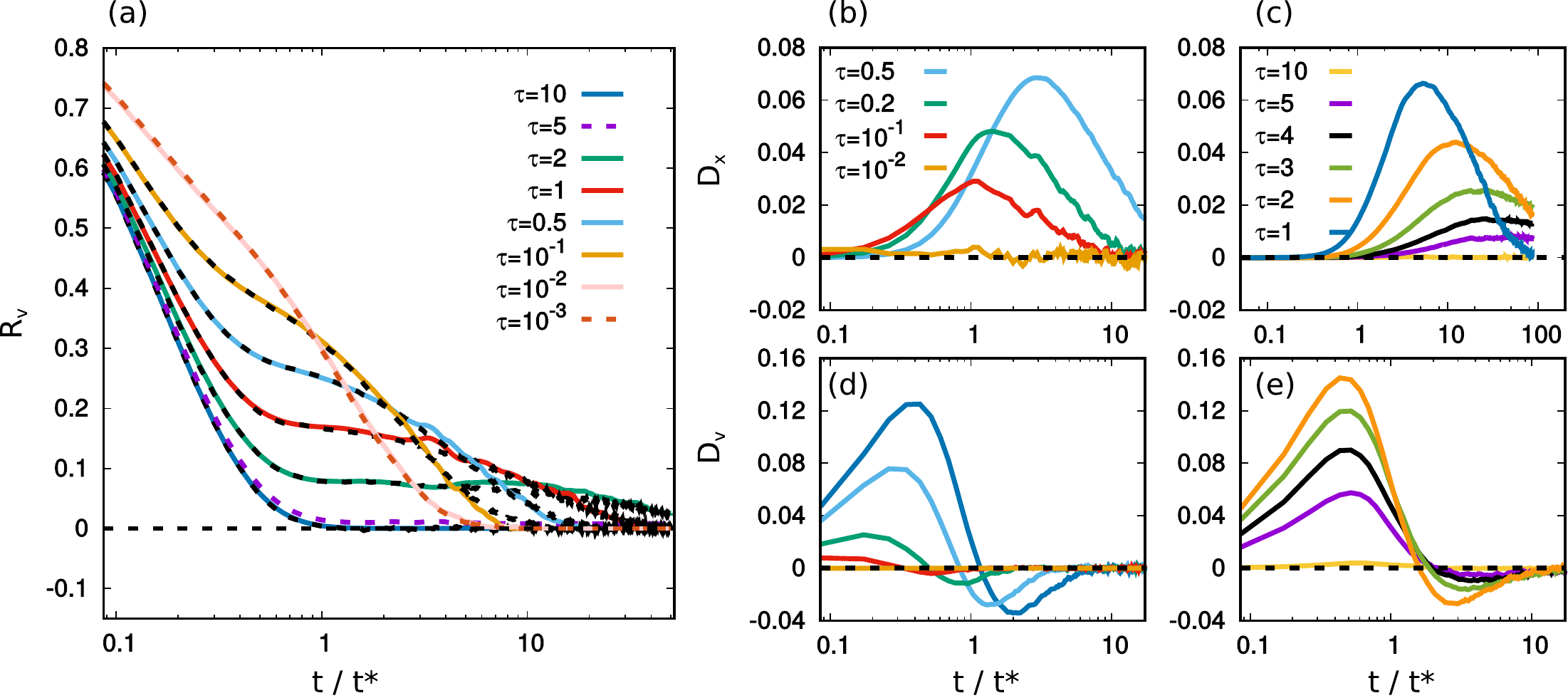}
\caption{\label{fig:Rdoublewell}
Panel (a): Response function, $R_{x}(t)$, for different values of $\tau$ calculated numerically via Eq.~\eqref{eq:numericalresponse} in the case of a double-well potential, $U(x)=k (x^4/4 - x^2/2)$.
The dashed black lines are obtained by using the FDR, Eq.~\eqref{eq:second_result}.
Panel (b) and (c): $\mathcal{D}_x(t)$ for different values of $\tau$ as reported in the legend.
Panel (d) and (e):  $\mathcal{D}_v(t)$ for different values of $\tau$ as reported in the legend.
The other parameters are $D_a=1$, $k=3$, $\gamma=1$.
}
\end{figure}

The response function, in the double-well potential case, $U(x)=k(x^4/4
- x^2/2)$, is reported in Fig.~\ref{fig:Rdoublewell}~(a) for different
values of $\tau$, exploring both the small and the large persistence
regimes.  In the small persistence regime, $\mathcal{R}_x(t)$ are collapsed onto
the same curve for a broad range of $\tau$ (for $\tau\leq10^{-2}$). 
As also occurs in the quartic potential case, the activity is the faster degree of freedom
and can be approximated by a white noise, $f_a \approx \sqrt{2D_a}
\xi$, so that $\tau$ does not play any role.  
In this case, the relaxation towards zero is mainly determined
by two time-regimes, as usual in the case of passive Brownian
particles.  The first is determined by the relaxation towards the
minimum of one of the two wells, the second accounts for the jump from
a well to another.  When $\tau$ is increased, the first time-regime
starts decreasing faster while the second time-regime decays more
slowly approaching zero for very larger times $t/t^* \gg \tau$.
Moreover, the increase of $\tau$ also reduces the role played by the second time-regime in the temporal decay of $\mathcal{R}_x(t)$ (the larger $\tau$, the smaller the value of $\mathcal{R}_x(t)$ when the second time regime starts occurring).
The second time-regime is even suppressed when $\tau$ is sufficiently large ($\tau \geq 5$, with the potential setting employed in Fig.~\ref{fig:Rdoublewell}~(a)).  
This suppression is consistent with the
result reported by Fily~\cite{fily2019self} in the infinite
persistence regime where the steady-state density distribution exactly
vanishes in the regions where the potential is concave (meaning that no jumps occur).
More generally, the statistical relevance of the second time-regime decreases with the increase of the persistence time
since the jumps from a minimum to the other occurs rarely when $\tau$ is increased both in the small $\tau$ regime~\cite{wio1989path, bray1990path, sharma2017escape} and in the large $\tau$ regime~\cite{woillez2020nonlocal}.

As occurs in the case of the quartic potential, the time-decay of
$\mathcal{R}_{x}(t)$ is dominated by the positional correlation or the velocity
correlations, in the small and large persistence regimes, respectively
(not shown).  In Fig.~\ref{fig:Rdoublewell}~(b), (c), (d) and (e),
we report $\mathcal{D}_x(t)$ and $\mathcal{D}_v(t)$ for different
values of $\tau$, revealing an interesting scenario. In analogy with
the quartic potential case, the $\mathcal{D}_x(t)$ assumes positive
values via a single peak that shifts for larger $t/t^*$ when $\tau$ is increased.
The $\mathcal{D}_v(t)$ displays an oscillation around
zero similarly to the quartic potential case.  
In a first range of $\tau$, the amplitudes of both
$\mathcal{D}_x(t)$ and $\mathcal{D}_v(t)$ increase with $\tau$ (as
shown in panels (b) and (d)).  A further increase of $\tau$ (panels~(d) and~(e)) produces
the amplitude decrease of both $\mathcal{D}_x(t)$ and
$\mathcal{D}_v(t)$, until both the functions become flat, approximatively for $\tau \geq 10$.  This
non-monotonic behavior with the persistence time implies that the system shows an
optimal value of $\tau$ that maximizes the departure from the
equilibrium via the breaking of the detailed balance.  We justify this
non-monotonic behavior through the phenomenology reported in~\cite{caprini2019activedoublewell}. 
When the particle moves close to one of the two minima, the particle explores a near-equilibrium regime where the local detailed balance almost holds and the local currents are almost zero both in the small and large persistence regimes. This explains why, in this space region, effective equilibrium approaches (that assume the detailed balance condition) give good predictions for the local steady-state properties.
Instead, when the particle overcomes the
inflected point of the potential (for which $\partial^2_x U =0$), the
particle enters into a non-equilibrium region that, in the large persistence regime, displays an effective
negative mobility.
There, the particle velocity increases
exponentially in time (with very small fluctuations) until the second
minimum is reached through an accelerated motion.  
Thus, most of the non-equilibrium currents are generated when the particle jumps.
As we already discussed, the number of jumps from a minimum to the other decreases with $\tau$ (in particular, until to be suppressed for $\tau=10$ and our setting of the potential).
Our measure of $D_x$ and $D_v$ confirms that the detailed balance is mostly broken when the jumps occur and, thus, it is reasonable to assume that the detailed balance is almost restored for $\tau$ large enough where no jumps occur.

\subsection{Measuring the non-equilibrium in active systems}
\label{measuring}

The study of the correlations and, in particular, the time behavior
of $\mathcal{D}_x(t)$ and $\mathcal{D}_v(t)$, suggests introducing a
measure to quantify the breaking of the detailed balance and, thus,
how much the system is far from equilibrium.  Since $\mathcal{D}_v(t)$
(and, in principle, also $\mathcal{D}_x(t)$) could assume negative
values, we need to consider $|\mathcal{D}_v(t)|$ and
$|\mathcal{D}_x(t)|$.  We propose a measure based on the time integral
of these observables and introduce the quantity:
\begin{equation}
\label{eq:def_mu}
\mu=\mu_x + \mu_v \,,
\end{equation}
where 
\begin{flalign}
\label{eq:def_mux}
\mu_x &= \frac{1}{t^*}\int_0^{\infty} |\mathcal{D}_x(s)| ds\\
\label{eq:def_muv}
\mu_v &= \frac{1}{t^*}\int_0^{\infty} |\mathcal{D}_v(s)| ds \,,
\end{flalign}
where $t^*$ is the typical time that rules the relaxation of the
response function for the passive case (see above Sec.~\ref{sec:numerics}).
The observable $\mu$ has the following meaning: starting from an
initial configuration (which is averaged), $\mu$ measures: i) how much
the system is far from the equilibrium during the whole trajectory
history; ii) the global impact of the detailed balance
breaking on the time-decay of the response function.
Indeed, when the detailed balance holds, both $\mu_x$ and $\mu_v$ vanish since $D_x(s)=D_v(s)=0$.
In general, the larger is $\mu$, the greater is the aumount and how long the detailed balance has been broken starting from an initial configuration.

\begin{figure}[!t]
\centering
\includegraphics[width=0.85\linewidth,keepaspectratio]{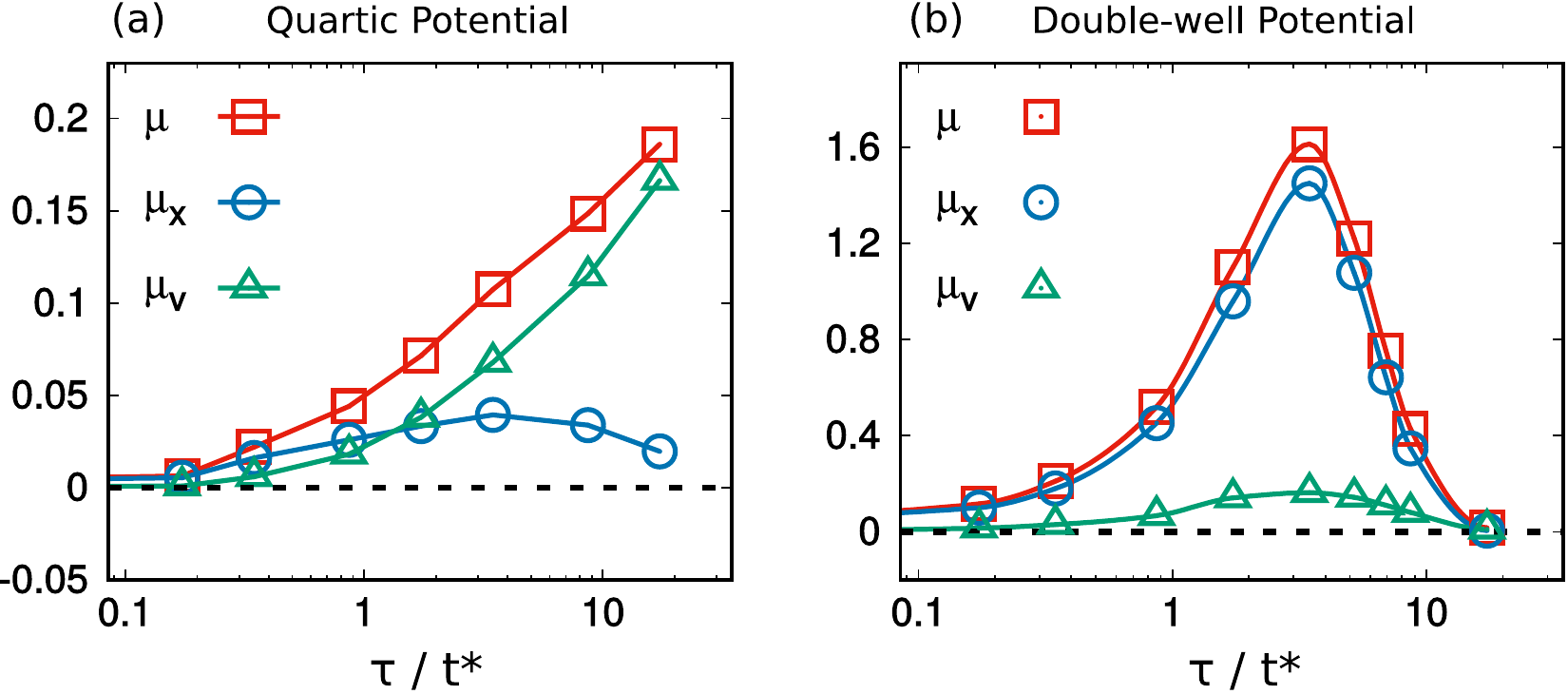}
\caption{\label{fig:mu}
$\mu$, $\mu_x$ and $\mu_v$ defined in Eqs.~\eqref{eq:def_mu},~\eqref{eq:def_mux} and~\eqref{eq:def_muv}, as a function of $\tau/t^*$ for the quartic potential, $U(x)=kx^4/4$, and the doublewell potential, $U(x)=k(x^4/4-x^2/2)$ for panels~(a) and~(b), respectively.
The other parameters are $D_a=1$, $k=3$, $\gamma=1$. 
}
\end{figure}

These observables are shown in Fig.~\ref{fig:mu}~(a) and~(b) for the
quartic and double-well potentials, respectively, for different values of $\tau/t^*$.  
In the quartic potential case, $\mu$ is an increasing function of $\tau$ meaning that
the persistence of the activity increases the departure from the equilibrium.  While $\mu \approx \mu_x$ in the
small persistence regime, this is no longer true
in the large persistence regime where $\mu \approx \mu_v$.  The
double-well potential case confirms the non-monotonic behavior already
observed in Fig.~\ref{fig:Rdoublewell}~(b)-(e): $\mu$ shows a peak
and then starts decreasing for larger values of
$\tau$ until to become almost zero.  Moreover, $\mu$ is mainly
determined by $\mu_x$ for the whole set of $\tau$ values.  Indeed,
even if $D_v(t)$ assumes also larger value than $D_x(t)$, it is
different from zero just for a narrow initial time-window while
$D_x(t)$ remains larger for a longer time.

\section{Failure of the aproximated approaches}\label{Sec:V}

To provide a qualitative explanation of several phenomena typical of
active matter, such as particle accumulation near boundaries or
the dynamics in confining potentials, different approximation schemes
have been successfully introduced.  Among the
others, the Unified Colored Noise approximation (UCNA) has been crucial for theoretical purposes, providing
the first analytical results for the probability distribution
function of both interacting and non-interacting confined systems \cite{maggi2015multidimensional}.
This approach provides a scheme to replace the
self-propulsion with effective interactions and is quite similar to the so-called Fox approximation~\cite{wittmann2017effective}. 
Moreover, it is derived assuming vanishing currents and, thus, the detailed
balance, providing the best equilibrium-like predictions to describe
the AOUP non-equilibrium dynamics.

Despite these approaches are very useful to understand the static properties of self-propelled particles, at least when confined through convex potentials or interacting through convex interactions, it has been shown that they fail to describe the active time-dependent properties, such as time-correlations and response functions. This concept has been already stressed in~\cite{caprini2018linear} where the authors show that the UCNA approximation is able to reproduce the response function just in small persistence regimes, where the system is near the equilibrium and a generalized FDR could be obtained perturbatively in $\tau$ using Eq.~\eqref{eq:response_vulpio}. In the large persistence regime, the numerical study for non-linear potentials shows the failure of the UCNA approach.

The aim of this Section is to enforce this idea, through additional
analytical arguments, showing that the breakdown of the detailed
balance plays an important role in correlations and response
functions and, as a consequence, also for susceptibility and effective
temperature.

\begin{figure}[!t]
\centering
\includegraphics[width=0.95\linewidth,keepaspectratio]{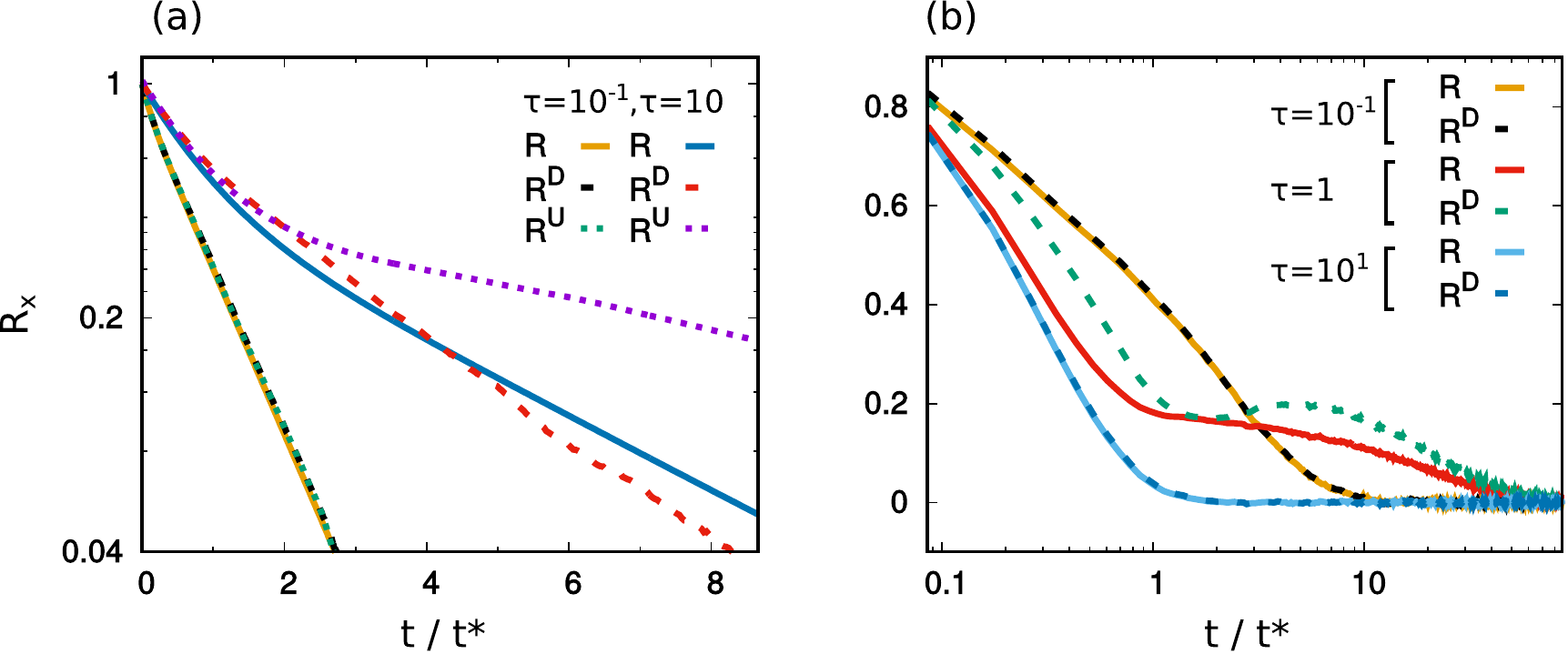}
\caption{\label{fig:responsevstime}
Response function, $R_{x}(t)$ as a function of $t/t^*$ for different values of $\tau$ as indicated in the legend, in the case of a quartic potential, $U(x)=k x^4$, and a double-well potential, $U(x)=k (x^4/4 - x^2/2)$, shown in panels (a) and (b), respectively.
Solid lines are calculated numerically via Eq.~\eqref{eq:numericalresponse}. Dashed lines report the expressions for $R^D$ (Eq.~\eqref{eq:RDX}) while dotted lines those for $R^U$ (Eq.~\eqref{eq:response_UCNApred}).
The parameters of the simulations are $D_a=1$, $k=3$, $\gamma=1$.
}
\end{figure}

\subsection{Assuming the detailed balance}

A possible approximated approach could consist in assuming the detailed balance condition in Eq.~\eqref{eq:second_result}.
This choice means that $\mathcal{D}_x(t)=0$ and $\mathcal{D}_v(t)=0$ and should lead to results similar to those obtained in the UCNA approximation.
Thus, the following relations hold:
 \begin{flalign}
&\left\langle x_i(t) \nabla_j U(0) \right\rangle =\left\langle \nabla_i U(t) x_j(0) \right\rangle\\
&\left\langle v_k(t) \nabla_k \nabla_i U(t) v_j(0) \right\rangle =\left\langle v_i(t) \nabla_i \nabla_k U(0) v_j(0) \right\rangle \,,
\end{flalign}
in such a way that the response can be approximated by
$$
\mathcal{R}^D_{x_ix_j}(t) \approx \frac{1}{D_a} \left\langle x_i(t) \frac{\nabla_j U(0)}{\gamma} \right\rangle + \frac{\tau^2}{D_a} \left\langle v_k(t) \frac{\nabla_k \nabla_i U(t)}{\gamma} v_j(0) \right\rangle \,,
$$
where the superscript $D$ stands for detailed balance.
Using the property $\langle v_k(t) \nabla_k \nabla_iU(t) v_j(s) \rangle=- d/dt \langle v_k(t) \nabla_k \nabla_i U(t) x_j(s) \rangle$, applying the derivative and using the equation of motion, we obtain
\begin{equation}
\label{eq:RDX}
\begin{aligned}
\mathcal{R}^D_{x_i x_j}(t) \approx &\frac{1}{D_a} \left\langle x_i(t) \frac{\nabla_j U(0)}{\gamma} \right\rangle + \frac{\tau}{D_a} \left\langle\frac{x_i(t)}{\gamma} \frac{\nabla_i \nabla_k U(0)}{\gamma} \nabla_k U(0) \right\rangle\\
&- \frac{\tau}{D_a} \left\langle x_i(t) \Gamma_{jk}(0) \frac{\nabla_k \nabla_n U(0)}{\gamma} v_n(0) \right\rangle -\frac{\tau^2}{D_a} \left\langle x_i(t) \frac{\nabla_j \nabla_k \nabla_n U(0)}{\gamma} v_k(0) v_n(0) \right\rangle \,,
\end{aligned}
\end{equation}
where we have also used the reversibility condition holding because of
the detailed balance assumption. The results of this approximation are
shown in Fig.~\ref{fig:responsevstime} for the quartic potential and
the double-well potential. As expected, in the case of a quartic
potential the approximation holds for small $\tau=10^{-1}$, i.e. in near equilibrium regimes,
while is less accurate for large $\tau=10$. In the case of a double-well
potential, the approximation holds for small and large values of
$\tau$, while is not accurate in the intermediate regime.
We remark that the failure of $\mathcal{R}^D_x(t)$ to reproduce the decay of $\mathcal{R}_x(t)$ is much evident 
in correspondence of the ranges of $\tau$ such that the measure of the non-equilibrium proposed in this work, Eq.~\eqref{eq:def_mu}, assumes large values.

We also stress that $R^D_{ij}(t)$ is expressed in the following form:
$$
D_a \mathcal{R}^D_{x_i x_j}(t)\propto \langle x_i(t) \,C_j(0) \rangle \,,
$$
where $C_j(0)$ is defined by Eq.\eqref{eq:RDX} that apparently
looks like similar to Eq.\eqref{eq:response_vulpio}.  Moreover, we also
point out that it is not possible to express $C_j$ as $\propto d/dx_j
\,\nu(\mathbf{x}, \mathbf{v})$ if we require that $\nu$ is a
normalized probability as in Eq.\eqref{eq:response_vulpio}.  This
observation alone implies that any approximation for the probability
distribution could not agree with the detailed balance assumption.


\subsection{Unified Colored Noise approximation}

According to the UCNA scheme~\cite{maggi2015multidimensional,
  marconi2015towards, caprini2019activity}, we can explicitly derive
an approximate solution for the steady-state probability distribution
holding for general potentials that has been used to reproduce the accumulation near boundaries~\cite{maggi2015multidimensional} (modeled as soft truncated
potentials) and the particle accumulation far from the potential minimum of a single-well potential~\cite{caprini2019activity}.
Moreover, as shown in~\cite{caprini2018linear}, this procedure obtained neglecting the
particle velocity leads to wrong results for time-correlations and
response functions even in the harmonic case (that can be analytically
checked).  Recently, the UCNA has been successively
extended to include the particle velocity~\cite{marconi2016velocity,
  caprini2019activity}, that displays a Gaussian-like shape with a
space-dependent velocity variance.  The extended version of the UCNA
has the following form:
\begin{equation}
\label{eq:approximate_pxv}
p(\mathbf{x},\mathbf{v}) \propto  \exp{\left( - \frac{\tau}{D_a} \mathbf{v} \frac{\boldsymbol{\Gamma}(\mathbf{x})}{2} \mathbf{v}\right)} \exp{\left( - \frac{\mathcal{H}}{D_a\gamma} \right)} \,,
\end{equation}
where the effective Hamiltonian $\mathcal{H}$ and $\boldsymbol{\Gamma}(\mathbf{x})$ are given by:
\begin{flalign}
&\mathcal{H}(\mathbf{x})= U(\mathbf{x}) + \frac{\tau}{2\gamma} [U'(\mathbf{x})]^2 - D_a \gamma \log{\left(\text{det}\boldsymbol{\Gamma}(\mathbf{x})\right)} \\
&\boldsymbol{\Gamma}(\mathbf{x})=\mathcal{I}+ \frac{\tau}{\gamma} \nabla^2U(\mathbf{x}) \,.
\end{flalign}
Here, the matrix $\boldsymbol{\Gamma}(\mathbf{x})$ plays the role of a space-dependent Stokes force that affects both the effective potential and the variance of the particle velocity. We remind that these interpretations should be limited to the convex potential (or non-convex potential in the small persistence regime) since the approximation cannot work if $\boldsymbol{\Gamma}(\mathbf{x})$ assumes non-positive values in some regions of space. Thus, there is no hope of applying the method in the case of double-well potentials.

Using the UCNA, it is possible to derive an expression for the linear response function, $\mathcal{R}_x(t)$. Indeed, plugging the log-derivative of Eq.~\eqref{eq:approximate_pxv} into Eq.~\eqref{eq:response_vulpio}, we obtain:
\begin{equation}
\label{eq:response_UCNApred}
\begin{aligned}
\mathcal{R}^U_{ij}(t) &\approx \frac{1}{D_a} \left\langle x_i(t) \frac{\nabla_j U(0)}{\gamma} \right\rangle + \frac{\tau}{D_a} \left\langle\frac{x_i(t)}{\gamma} \frac{\nabla_j \nabla_k U(0)}{\gamma} \nabla_k U(0) \right\rangle\\
&- \frac{\tau}{D_a} \left\langle x_i(t)  \Gamma_{jk}(0) \frac{\nabla_k \nabla_n U(0)}{\gamma} v_n(0) \right\rangle -\frac{\tau^2}{D_a} \left\langle x_i(t) \frac{\nabla_j\nabla_n \nabla_k U(0)}{\gamma} \left( \frac{3}{2}\frac{D_a}{\tau}\Gamma^{-1}_{nk}(0)  -\frac{K_{nk}(0)}{2}  \right) \right\rangle \,,
\end{aligned}
\end{equation}
where the superscript $U$ stands for UCNA and $K_{ij}=v_j(s)v_k(s)$ is
the generalized kinetic energy.  The derivation of this result is
reported in Appendix~\ref{App:UCNA}.  The results of the UCNA
approximation are shown in Fig.~\ref{fig:responsevstime}(a) for the
quartic potential.
$\mathcal{R}^U_x(t)$ is a good approximation of $\mathcal{R}_x(t)$ just in the small persistence regime (here, shown for $\tau=10^{-1}$), while is
not accurate for large $\tau=10$ since $\mathcal{R}^U_x(t)$ decays slower than $\mathcal{R}_x(t)$.

We stress that $\mathcal{R}^U_x(t)$ in general does not coincide with $\mathcal{R}^D_x(t)$ because the
last terms in the second lines of Eq.~\eqref{eq:response_UCNApred}
and Eq.~\eqref{eq:RDX} are different.  However, according to the
UCNA prediction, $\tau \Gamma_{kj}(\mathbf{x}) \langle K_{kj}
(\mathbf{x}) \rangle \approx D_a$, where $\langle K_{ij} (x) \rangle$
is the space-dependent kinetic temperature introduced in
Ref.~\cite{marconi2017heat, caprini2019activity} for the scalar case (and the last average is realized at fixed $x$).
Eq.~\eqref{eq:response_UCNApred} coincides with Eq.~\eqref{eq:RDX} only by using this reasonable identification.  
We also stress that the UCNA dramatically fails to describe active
particles in non-convex potentials because the matrix
$\boldsymbol{\Gamma}^{-1}$ is ill-defined.




\section{Conclusion}
\label{sec:concl}

Our study provides an explicit extension of the generalized FDR to
non-equilibrium systems of active matter, that are numerically checked
in many cases of interest, in particular, the case of the quartic
potential explored for one, two and three dimensional systems and the
case of a double-well potential.  At variance with previous
approaches, we derive simple expressions for the correlators involved
in the FDR, which could be measured in experimental systems.
Our results can be extended to more general dynamics, within the
framework of non-equilibrium systems.


Through our expression, we are also able to extrapolate a measure to
determine how much the system is far from equilibrium evaluating its
influence on the time-decay of the response function.  We quantify how
much the breaking of the detailed balance affects the response decay
looking directly at the correlations of the generalized FDR.  This
analysis agrees with previous qualitative observations showing a
monotonic increase of the departure from the equilibrium with the
increase of the persistence time in the quartic potential case but a
reentrant (non-monotonic) behavior in the case of a double-well
potential where the equilibrium is almost restored for large enough persistence times.  
Finally, as the intuition suggests, increasing the dimensionality
reduces the departure from equilibrium.

Our expression of the FDR shows that the breaking of the
detailed balance plays a crucial role to determine the time-decay of
the response function and, thus, that any approximated descriptions
that assume the detailed balance (or similar hypothesis to neglect the
dynamical properties of active particles) to calculate responses,
susceptibilities or mobilities are expected to yield poor
results, unless in regimes of small activity, or when interactions
are weak and confinement is absent or harmonic (all cases where
detailed balance is weakly or not violated by this model).


\vspace{6pt}

\section{Acknowledgements}

This research was funded by MIUR PRIN 2017 grant number 201798CZLJ and by Regione
Lazio through the Grant ”Progetti Gruppi di Ricerca”
N. 85-2017-15257.
The authors thank A. Vulpiani for useful discussions.


\appendix

\section{Derivation of the FDR for active particles}{\label{App:Novikov}}

As a first step, we manipulate the original dynamics (i.e. Eqs.~\eqref{eq:motion} and~\eqref{eq:motion2}) to eliminate $\mathbf{f}_a$ in favor of $\dot{\mathbf{x}}=\mathbf{v}$, through an exact change of variables.
Applying the time-derivative to Eq.~\eqref{eq:motion}, using Eq.\eqref{eq:motion2} to eliminate $\dot{\mathbf{f}}_a$ and, finally, replacing $\mathbf{f}_a$ with $\mathbf{v}$ (through Eq.~\eqref{eq:motion}), we get:
\begin{flalign}
\dot{\mathbf{x}}&=\mathbf{v}\\
\label{eq:different_dynamics}
\tau\dot{\mathbf{v}}&=- \boldsymbol{\Gamma}(\mathbf{x}) \mathbf{v} - \frac{\nabla U(\mathbf{x})}{\gamma} + \boldsymbol{\chi} + \frac{{\mathbf{H}}^t}{\gamma} \,, 
\end{flalign}
where $\boldsymbol{\chi}=\sqrt{2D_a}\boldsymbol{\xi}$ is a noise vector such that 
$$
\langle \chi_{i}(t) \chi_{j}(s)\rangle = 2D_a\delta_{ij}\delta(t-s) \,.
$$
The term $\boldsymbol{\Gamma}(\mathbf{x})$ is a two-dimensional matrix acting as a space-dependent friction whose components read:
$$
\Gamma_{ij}(\mathbf{x})=\delta_{ij}+\frac{\tau}{\gamma} \nabla_i \nabla_j U(\mathbf{x}) \,,
$$ 
where the space-dependence is provided by the potential curvature.
The original dynamics, perturbed on the particle's position,
is mapped onto an underdamped dynamics with an effective perturbation that is not simply $\mathbf{h}^t$ as in Eqs.~\eqref{eq:motion}.
The dynamics of $\mathbf{v}_i$ is affected by an effective perturbation ${{\mathbf{H}}}^t$ that reads:
\begin{equation}
\frac{{\mathbf{H}}^t}{\gamma} =  \left[1 + \tau\frac{d}{dt}\right] \frac{\mathbf{h}^t}{\gamma}
  \end{equation}
Despite the strangeness of the transformed dynamics, it is straightforward to apply the Novikon theorem~\cite{n65} to get an exact expression for $\mathcal{R}_{x_i x_j}(t-s)$ in terms of noise correlations. Indeed, since the functional derivative with respect to $\boldsymbol{\mathcal{H}}^t/\gamma$ is equivalent to the functional derivative with respect to the noise $\boldsymbol{\chi}$:
$$
 \frac{\delta}{\delta \chi_j} =\gamma \frac{\delta}{\delta H}_j =\left[1 + \tau\frac{d}{dt}\right] \gamma \frac{\delta}{\delta h_j(t)}\,,
$$
where we have used the chain rule in the last equality. 
The inversion of this relation reads:
\begin{equation}
\label{eq:change_var_hchi}
\gamma \frac{\delta}{\delta h_j(t)}=\left[1 - \tau\frac{d}{dt}\right] \frac{\delta}{\delta \chi_j(t)}\,.
\end{equation}
Thus, starting from the definition of $\mathcal{R}_{x_ix_j}(t-s)$, with $t>s$, we get:
\begin{equation}
\begin{aligned}
\mathcal{R}_{x_ix_j}(t-s) &= \frac{\langle x_i(t)\rangle^h-\langle x_i(t)\rangle}{\delta x_j(s)}=\gamma\left.\frac{\delta\langle x_i(t)\rangle^h}{\delta h_j(s)}\right|_{h=0}=\gamma\int^t  \mathcal{D}[\boldsymbol{\chi}_h] P[\boldsymbol{\chi}_h] \frac{\delta}{\delta h_j(s)} x^h_i(t) \\
&= -\gamma\int^t  \mathcal{D}[\boldsymbol{\chi}_h] x^h_i(t)\frac{\delta}{\delta h_j(s)} P[\boldsymbol{\chi}_h] = -\int^t  \mathcal{D}[\boldsymbol{\chi}_h] x^h_i(t)\left[1 - \tau\frac{d}{ds}\right] \frac{\delta}{\delta \chi_j(s)} P[\boldsymbol{\chi}_h] \,,
\end{aligned}
\end{equation}
where, we have used the integration by parts in the third equality and Eq.~\eqref{eq:change_var_hchi} in the last equality.
The term $P[\boldsymbol{\chi}_h]$ is the noise path probability, generating the trajectory, namely:
$$
P[\boldsymbol{\chi}_h] \propto \exp{\left(-\frac{1}{4D_a}\int^s \boldsymbol{\chi}_h^2(t') dt' \right)} \,,
$$
where the index $h$ reminds that $\boldsymbol{\chi}_h$ generates the perturbed trajectory.
Using the Gaussianity of $P[\boldsymbol{\chi}_h]$ and integrating by parts, we get:
\begin{equation}
\label{eq:response_harmonic_thirdapproach}
{2D_a}\mathcal{R}_{i j}=\langle x_i(t) \chi_j(s) \rangle - \tau\frac{d}{ds}  \langle x_i(t) \chi_j(s) \rangle \,.
\end{equation}
This expression is a generalization to the active dynamics, employed so
far, of the well-known FDR for overdamped passive Brownian particles,
which can be recovered in the equilibrium limit, $\tau\to0$.  The
additional term, which is proportional to $\tau$ involves the time
derivative of the noise correlation and needs to be treated carefully.
Replacing the noise term $\chi_i$ with Eq.~\eqref{eq:different_dynamics}, that is the evolution equation for $v_i$, in Eq.~\eqref{eq:response_harmonic_thirdapproach}, the correlation between position and noise reads:
\begin{equation}
\langle x_i(t) \chi_j(s) \rangle = \tau \frac{d}{ds} \langle x_i(t) v_j(s) \rangle +\frac{\tau}{\gamma}\langle x_i(t) v_j(s) \rangle +\langle x_i(t) \nabla_{i}\nabla_k v_k(s) \rangle  + \frac{1}{\gamma}\langle x_i(t) \nabla_j U(s) \rangle \,.
\end{equation}
Plugging this expression into Eq.~\eqref{eq:response_harmonic_thirdapproach}, it is straightforward to derive an expression for $R_{ij}(t-s)$ as a function of more suitable correlations involving particle position and velocity:
\begin{equation}
\label{eq:maes_selfprop}
\begin{aligned}
{2D_a} \mathcal{R}_{ij}(t-s) =  \left\langle  x_i(t) \left[ v_j(s) + \frac{\nabla_j U(s)}{\gamma}  \right] \right\rangle  - \tau^2\frac{d^2}{ds^2}\left\langle  x_i(t)  v_j(s)  \right\rangle - \tau^2\frac{d}{ds}\left\langle  x_i(t)   \frac{\nabla_j\nabla_k U(s)}{\gamma} v_k(s)  \right\rangle \,.
\end{aligned}
\end{equation}
The first average corresponds to the well-known passive Brownian result, while the second and the third averages account for the non-equilibrium contributions occurring for non-vanishing $\tau$.

As it is, expression~\eqref{eq:maes_selfprop} is not so useful because involves the temporal derivatives of suitable correlation functions and cannot be easily calculated.
Using the steady-state property of the correlations, that implies that $d/ds \to - d/dt$, we get:
\begin{equation}
\label{eq:for_thesusceptibility}
\begin{aligned}
{2D_a}\mathcal{R}_{ij}(t-s) &= \left\langle  x_i(t) \left[ v_j(s) + \frac{\nabla_j U(s)}{\gamma}  \right] \right\rangle  - \tau^2\frac{d}{dt}\left\langle  v_i(t)  v_j(s)  \right\rangle + \tau^2\left\langle  v_i(t)   \frac{\nabla_j \nabla_k U(s)}{\gamma} v_k(s)  \right\rangle \,.
\end{aligned}
\end{equation}
Now, replacing $\dot{v}$ with the equation of motion, Eq.~\eqref{eq:different_dynamics}, we obtain:
\begin{equation}
\begin{aligned}
{2D_a} \mathcal{R}_{ij}(t-s) =& \left\langle  x_i(t) \left[ v_j(s) + \frac{\nabla_j U(s)}{\gamma}  \right] \right\rangle  + \tau\left\langle  \Gamma_{ik}(t) v_k(t)  v_j(s)  \right\rangle + \tau\left\langle  \frac{\nabla_i U(t)}{\gamma}  v_j(s)  \right\rangle \\
&+ \tau^2\left\langle  v_i(t)   \frac{\nabla_j \nabla_k U(s)}{\gamma} v_k(s)  \right\rangle \,,
\end{aligned}
\end{equation}
where we have used the causality, such that $\langle \chi_i(t) O_j(s)\rangle=0$, for any observable $O_j$ if $t>s$.
Using the same strategy, further manipulations lead to the final result:
\begin{equation}
\begin{aligned}
{2D_a}\mathcal{R}_{x_ix_j}(t-s) =&
\left[ \left\langle x_i(t) \frac{\nabla_j U(s)}{\gamma} \right\rangle +\left\langle \frac{\nabla_i U(t)}{\gamma} x_j(s) \right\rangle  \right]\\
&+\tau^2\left[ \left\langle v_k(t) \frac{\nabla_k \nabla_i U(t)}{\gamma} v_j(s) \right\rangle +\left\langle v_i(t) \frac{\nabla_j\nabla_k U(s)}{\gamma} v_k(s) \right\rangle  \right] \,,\\
\end{aligned} 
\end{equation}
where, again, we have replaced $\dot{v}_i$ with the equation of motion
ad used that $\langle \chi_i(t) O_j(s)\rangle=0$, for any observable
$O_j$ if $t>s$.  This completes the derivation of
Eq.\eqref{eq:second_result}.  Since the system does not satisfy the
detailed balance we cannot use the reversibility to simplify the above
expression, because $\left\langle x_i(t) \nabla_j U(s) \right\rangle
\neq\left\langle \nabla_i U(t) x_j(s) \right\rangle$ and $\left\langle
v_k(t) \nabla_k \nabla_i U(t) v_j(s) \right\rangle \neq\left\langle
v_i(t) \nabla_j \nabla_k U(s) v_k(s) \right\rangle$.


\section{The active harmonic oscillator}{\label{App:Harmonioscillator}}

The harmonic oscillator can be used as a check to test the different FDR for the linear response function, namely Eq.~\eqref{eq:response_vulpio} and Eq.~\eqref{eq:second_result}.
Indeed, in this case, the response function and the correlations can be calculated exactly as a function of time and the steady-state distribution is known.
Since the system is linear, each component of the dynamics evolves independently with the others and, thus, studying the one-dimensional system is enough. For this reason, we drop the subscripts in what follows.
Choosing $U(x)=k/2 x^2$, the steady-state distribution is a multivariate Gaussian, namely:
\begin{equation}
p(x,v) \propto \exp{\left(-\frac{\tau}{D_a} \Gamma \frac{v^2}{2}   \right)} \exp{\left(- \frac{k}{D_a\gamma}\Gamma \frac{x^2}{2} \right)} 
\end{equation}
where $v=\dot{x}$ corresponds to the particle's velocity, such that:
\begin{equation}
\label{eq:linear_change_var}
v=-\frac{k}{\gamma} x + \frac{\text{f}_a}{\gamma} \,,
\end{equation}
 and the coefficient $\Gamma$ is given by
$$
\Gamma=1+\tau\frac{k}{\gamma}\,,
$$
and represents the effective viscosity of the dynamics which is spatial independent in the harmonic case.
We start evaluating Eq.~\eqref{eq:response_vulpio}, that explicitly leads to the following expression for the response function in terms of correlations: 
\begin{equation}
\label{eq:Vulpianiformula}
\begin{aligned}
\mathcal{R}_x(t) &= - \left\langle x(t) \frac{d}{d x} p(x(0), v[x(0), f^a(0)])\right\rangle \\
&= \frac{k}{D_a \gamma} \Gamma \left[\left\langle x(t) x(0) \right\rangle - \tau \left\langle x(t) v(0) \right\rangle \right] \,,
\end{aligned}
\end{equation}
where, we $d/dx$ is the total derivative (not just the partial derivative) and the function needs to evaluated as a function of $x$ and $\text{f}_a$ because of Eq.~\eqref{eq:linear_change_var}.
We observe that $R_x$ reduces to the well-known relation for passive overdamped systems in the limit $\tau \to 0$, since in this limit $\Gamma \to 1$.
In this special case where the detailed balance holds, the correlations can be calculated as a function of time, in such a way that, in the steady-state, we get:
\begin{flalign}
\langle x(t) x(0)\rangle &= \frac{D_a\gamma}{k} \tau \left( \tau^2 \frac{k^2}{\gamma^2}-1\right)^{-1} \left[ \frac{k}{\gamma} e^{  -\frac{t}{\tau}} - \frac{1}{\tau} e^{  -\frac{k}{\gamma}t  } \right] \\
\langle v(t) x(0)\rangle &= \frac{d}{dt} \langle x(t) x(0)\rangle = -\langle x(t) v(0)\rangle \,.
\end{flalign}
Combining these results, we obtain
\begin{equation}
\label{eq:harmonic_hexactexpression_linear}
\begin{aligned}
\mathcal{R}_x(t) = e^{-\frac{k}{\gamma}t} \,,
\end{aligned}
\end{equation}
which does not depend on $\tau$.
As a consequence, the response function in the presence of a harmonic force is not influenced by the value of $\tau$.

In this special case, also the generalized FDR, given by Eq.\eqref{eq:second_result}, can be evaluated explicitly and reads:
\begin{equation}
D_a\gamma \mathcal{R}_x(t) = k \left[ \left\langle x(t) x(0) \right\rangle + \tau^2 \left\langle v(t) v(0)\right\rangle \right] \,.
\end{equation}
Using the time-dependent expressions for the correlation functions, we obtain Eq.~\eqref{eq:harmonic_hexactexpression_linear}, confirming that the equivalence between Eq.~\eqref{eq:response_vulpio} and Eq.~\eqref{eq:second_result} in the harmonic case.

\section{Response function with the UCNA approach}{\label{App:UCNA}}

To evaluate the response function using the UCNA approach, it is enough to plug the log-derivative of the UCNA probability distribution, namely Eq.\eqref{eq:approximate_pxv}, into the generalized FDR \eqref{eq:response_vulpio} without employing any path integral techniques.
Explicitly, the log-derivative of the distribution \eqref{eq:approximate_pxv} reads:
\begin{equation}
\label{eq:app_logpxv}
\begin{aligned}
-\frac{d}{dx_j} \log p(\mathbf{x},\mathbf{v}) = \frac{1}{D_a\gamma}\frac{d}{dx_j} \mathcal{H}  + \frac{\tau}{2 D_a} \frac{d}{dx_j} \left(v_k(\mathbf{x})\Gamma_kn(\mathbf{x}) v_n(\mathbf{x})\right) \,.
\end{aligned}
\end{equation}
We remind that the spatial derivative comparing in Eq.\eqref{eq:response_vulpio} and, thus, in Eq.\eqref{eq:app_logpxv}, is a total derivative \cite{caprini2018linear}. 
As already shown in \cite{caprini2018linear}, one needs to replace the velocity $\mathbf{v}$ with its whole expression as a function of the position before taking the derivative, i.e. $\gamma\mathbf{v}=-\nabla U + \mathbf{f}_a$, otherwise one gets inconsistent results. Following these procedures, we get:
\begin{flalign}
&\frac{1}{D_a\gamma}\frac{d}{dx_j} \mathcal{H}(\mathbf{x})  = \frac{\nabla_k U}{D_a\gamma} \Gamma_{kj}(\mathbf{x}) -  \frac{3}{2} \frac{d}{dx_j} \log{\text{det} |\boldsymbol{\Gamma(\mathbf{x})}|}\\
&\frac{\tau}{2 D_a} \frac{d}{dx_j} \left(v_k(\mathbf{x}) \Gamma_{kn}(\mathbf{x}) v_n(\mathbf{x})\right) = \frac{1}{D_a} \frac{\tau^2}{2\gamma} \nabla_{j}\nabla_k \nabla_n U v_k v_n -\frac{\tau}{D_a} \Gamma_{jk}(\mathbf{x}) \frac{\nabla_k \nabla_n U(\mathbf{x})}{\gamma} v_n   \,.
\end{flalign}
The spatial log-derivative of the $\boldsymbol{\Gamma}$ determinant can be evaluated in components and reads:
\begin{equation}
\frac{\partial}{\partial x_n} \log{\text{det} \boldsymbol{|\Gamma|}}%
= \sum_{jk} \Gamma_{jk}^{-1} \frac{\partial}{\partial x_n} \Gamma_{kj} = \sum_{jk} \Gamma^{-1}_{jk} \frac{\partial}{\partial x_j} \Gamma_{kn}
\end{equation}
Finally, collecting the results all together in Eq.\eqref{eq:response_vulpio}, we obtain Eq.\eqref{eq:response_UCNApred}.



\bibliographystyle{apsrev4-1}

\bibliography{Respbib}

\begin{thebibliography}{74}%
\makeatletter
\providecommand \@ifxundefined [1]{%
 \@ifx{#1\undefined}
}%
\providecommand \@ifnum [1]{%
 \ifnum #1\expandafter \@firstoftwo
 \else \expandafter \@secondoftwo
 \fi
}%
\providecommand \@ifx [1]{%
 \ifx #1\expandafter \@firstoftwo
 \else \expandafter \@secondoftwo
 \fi
}%
\providecommand \natexlab [1]{#1}%
\providecommand \enquote  [1]{``#1''}%
\providecommand \bibnamefont  [1]{#1}%
\providecommand \bibfnamefont [1]{#1}%
\providecommand \citenamefont [1]{#1}%
\providecommand \href@noop [0]{\@secondoftwo}%
\providecommand \href [0]{\begingroup \@sanitize@url \@href}%
\providecommand \@href[1]{\@@startlink{#1}\@@href}%
\providecommand \@@href[1]{\endgroup#1\@@endlink}%
\providecommand \@sanitize@url [0]{\catcode `\\12\catcode `\$12\catcode
  `\&12\catcode `\#12\catcode `\^12\catcode `\_12\catcode `\%12\relax}%
\providecommand \@@startlink[1]{}%
\providecommand \@@endlink[0]{}%
\providecommand \url  [0]{\begingroup\@sanitize@url \@url }%
\providecommand \@url [1]{\endgroup\@href {#1}{\urlprefix }}%
\providecommand \urlprefix  [0]{URL }%
\providecommand \Eprint [0]{\href }%
\providecommand \doibase [0]{http://dx.doi.org/}%
\providecommand \selectlanguage [0]{\@gobble}%
\providecommand \bibinfo  [0]{\@secondoftwo}%
\providecommand \bibfield  [0]{\@secondoftwo}%
\providecommand \translation [1]{[#1]}%
\providecommand \BibitemOpen [0]{}%
\providecommand \bibitemStop [0]{}%
\providecommand \bibitemNoStop [0]{.\EOS\space}%
\providecommand \EOS [0]{\spacefactor3000\relax}%
\providecommand \BibitemShut  [1]{\csname bibitem#1\endcsname}%
\let\auto@bib@innerbib\@empty
\bibitem [{\citenamefont {Onsager}(1931)}]{O31}%
  \BibitemOpen
  \bibfield  {author} {\bibinfo {author} {\bibfnamefont {L.}~\bibnamefont
  {Onsager}},\ }\href@noop {} {\bibfield  {journal} {\bibinfo  {journal} {Phys.
  Rev.}\ }\textbf {\bibinfo {volume} {37}},\ \bibinfo {pages} {405} (\bibinfo
  {year} {1931})}\BibitemShut {NoStop}%
\bibitem [{\citenamefont {Kubo}(1957)}]{K57}%
  \BibitemOpen
  \bibfield  {author} {\bibinfo {author} {\bibfnamefont {R.}~\bibnamefont
  {Kubo}},\ }\href@noop {} {\bibfield  {journal} {\bibinfo  {journal} {J. Phys.
  Soc. Japan}\ }\textbf {\bibinfo {volume} {12}},\ \bibinfo {pages} {570}
  (\bibinfo {year} {1957})}\BibitemShut {NoStop}%
\bibitem [{\citenamefont {Marconi}\ \emph {et~al.}(2008)\citenamefont
  {Marconi}, \citenamefont {Puglisi}, \citenamefont {Rondoni},\ and\
  \citenamefont {Vulpiani}}]{marconi2008fluctuation}%
  \BibitemOpen
  \bibfield  {author} {\bibinfo {author} {\bibfnamefont {U.~M.~B.}\
  \bibnamefont {Marconi}}, \bibinfo {author} {\bibfnamefont {A.}~\bibnamefont
  {Puglisi}}, \bibinfo {author} {\bibfnamefont {L.}~\bibnamefont {Rondoni}}, \
  and\ \bibinfo {author} {\bibfnamefont {A.}~\bibnamefont {Vulpiani}},\
  }\href@noop {} {\bibfield  {journal} {\bibinfo  {journal} {Physics Reports}\
  }\textbf {\bibinfo {volume} {461}},\ \bibinfo {pages} {111} (\bibinfo {year}
  {2008})}\BibitemShut {NoStop}%
\bibitem [{\citenamefont {Cugliandolo}(2011)}]{cugliandolo2011effective}%
  \BibitemOpen
  \bibfield  {author} {\bibinfo {author} {\bibfnamefont {L.~F.}\ \bibnamefont
  {Cugliandolo}},\ }\href@noop {} {\bibfield  {journal} {\bibinfo  {journal}
  {Journal of Physics A: Mathematical and Theoretical}\ }\textbf {\bibinfo
  {volume} {44}},\ \bibinfo {pages} {483001} (\bibinfo {year}
  {2011})}\BibitemShut {NoStop}%
\bibitem [{\citenamefont {Puglisi}\ \emph {et~al.}(2017)\citenamefont
  {Puglisi}, \citenamefont {Sarracino},\ and\ \citenamefont
  {Vulpiani}}]{puglisi2017temperature}%
  \BibitemOpen
  \bibfield  {author} {\bibinfo {author} {\bibfnamefont {A.}~\bibnamefont
  {Puglisi}}, \bibinfo {author} {\bibfnamefont {A.}~\bibnamefont {Sarracino}},
  \ and\ \bibinfo {author} {\bibfnamefont {A.}~\bibnamefont {Vulpiani}},\
  }\href@noop {} {\bibfield  {journal} {\bibinfo  {journal} {Physics Reports}\
  }\textbf {\bibinfo {volume} {709}},\ \bibinfo {pages} {1} (\bibinfo {year}
  {2017})}\BibitemShut {NoStop}%
\bibitem [{\citenamefont {Agarwal}(1972)}]{A72}%
  \BibitemOpen
  \bibfield  {author} {\bibinfo {author} {\bibfnamefont {G.~S.}\ \bibnamefont
  {Agarwal}},\ }\href@noop {} {\bibfield  {journal} {\bibinfo  {journal} {Z.
  Physik}\ }\textbf {\bibinfo {volume} {252}},\ \bibinfo {pages} {25} (\bibinfo
  {year} {1972})}\BibitemShut {NoStop}%
\bibitem [{\citenamefont {Falcioni}\ \emph {et~al.}(1990)\citenamefont
  {Falcioni}, \citenamefont {Isola},\ and\ \citenamefont {Vulpiani}}]{FIV90}%
  \BibitemOpen
  \bibfield  {author} {\bibinfo {author} {\bibfnamefont {M.}~\bibnamefont
  {Falcioni}}, \bibinfo {author} {\bibfnamefont {S.}~\bibnamefont {Isola}}, \
  and\ \bibinfo {author} {\bibfnamefont {A.}~\bibnamefont {Vulpiani}},\
  }\href@noop {} {\bibfield  {journal} {\bibinfo  {journal} {Physics {L}etters
  {A}}\ }\textbf {\bibinfo {volume} {144}},\ \bibinfo {pages} {341} (\bibinfo
  {year} {1990})}\BibitemShut {NoStop}%
\bibitem [{\citenamefont {Gnoli}\ \emph {et~al.}(2014)\citenamefont {Gnoli},
  \citenamefont {Puglisi}, \citenamefont {Sarracino},\ and\ \citenamefont
  {Vulpiani}}]{gnoli2014}%
  \BibitemOpen
  \bibfield  {author} {\bibinfo {author} {\bibfnamefont {A.}~\bibnamefont
  {Gnoli}}, \bibinfo {author} {\bibfnamefont {A.}~\bibnamefont {Puglisi}},
  \bibinfo {author} {\bibfnamefont {A.}~\bibnamefont {Sarracino}}, \ and\
  \bibinfo {author} {\bibfnamefont {A.}~\bibnamefont {Vulpiani}},\ }\href@noop
  {} {\bibfield  {journal} {\bibinfo  {journal} {PLoS ONE}\ }\textbf {\bibinfo
  {volume} {9}},\ \bibinfo {pages} {e93720} (\bibinfo {year}
  {2014})}\BibitemShut {NoStop}%
\bibitem [{\citenamefont {Speck}\ and\ \citenamefont {Seifert}(2006)}]{ss06}%
  \BibitemOpen
  \bibfield  {author} {\bibinfo {author} {\bibfnamefont {T.}~\bibnamefont
  {Speck}}\ and\ \bibinfo {author} {\bibfnamefont {U.}~\bibnamefont
  {Seifert}},\ }\href@noop {} {\bibfield  {journal} {\bibinfo  {journal}
  {Europhys. Lett.}\ }\textbf {\bibinfo {volume} {74}},\ \bibinfo {pages} {391}
  (\bibinfo {year} {2006})}\BibitemShut {NoStop}%
\bibitem [{\citenamefont {Seifert}\ and\ \citenamefont
  {Speck}(2010)}]{seifert2010fluctuation}%
  \BibitemOpen
  \bibfield  {author} {\bibinfo {author} {\bibfnamefont {U.}~\bibnamefont
  {Seifert}}\ and\ \bibinfo {author} {\bibfnamefont {T.}~\bibnamefont
  {Speck}},\ }\href@noop {} {\bibfield  {journal} {\bibinfo  {journal} {EPL
  (Europhysics Letters)}\ }\textbf {\bibinfo {volume} {89}},\ \bibinfo {pages}
  {10007} (\bibinfo {year} {2010})}\BibitemShut {NoStop}%
\bibitem [{\citenamefont {Warren}\ and\ \citenamefont
  {Allen}(2014)}]{malliavin}%
  \BibitemOpen
  \bibfield  {author} {\bibinfo {author} {\bibfnamefont {P.~B.}\ \bibnamefont
  {Warren}}\ and\ \bibinfo {author} {\bibfnamefont {R.~J.}\ \bibnamefont
  {Allen}},\ }\href@noop {} {\bibfield  {journal} {\bibinfo  {journal}
  {Entropy}\ }\textbf {\bibinfo {volume} {16}},\ \bibinfo {pages} {221}
  (\bibinfo {year} {2014})}\BibitemShut {NoStop}%
\bibitem [{\citenamefont {Novikov}(1965)}]{n65}%
  \BibitemOpen
  \bibfield  {author} {\bibinfo {author} {\bibfnamefont {E.~A.}\ \bibnamefont
  {Novikov}},\ }\href@noop {} {\bibfield  {journal} {\bibinfo  {journal}
  {Soviet Physcis-JETP}\ }\textbf {\bibinfo {volume} {20}},\ \bibinfo {pages}
  {1290} (\bibinfo {year} {1965})}\BibitemShut {NoStop}%
\bibitem [{\citenamefont {Cugliandolo}\ \emph {et~al.}(1994)\citenamefont
  {Cugliandolo}, \citenamefont {Kurchan},\ and\ \citenamefont
  {Parisi}}]{CKP94}%
  \BibitemOpen
  \bibfield  {author} {\bibinfo {author} {\bibfnamefont {L.~F.}\ \bibnamefont
  {Cugliandolo}}, \bibinfo {author} {\bibfnamefont {J.}~\bibnamefont
  {Kurchan}}, \ and\ \bibinfo {author} {\bibfnamefont {G.}~\bibnamefont
  {Parisi}},\ }\href@noop {} {\bibfield  {journal} {\bibinfo  {journal} {J.
  Phys. I France}\ }\textbf {\bibinfo {volume} {4}},\ \bibinfo {pages} {1641}
  (\bibinfo {year} {1994})}\BibitemShut {NoStop}%
\bibitem [{\citenamefont {Baiesi}\ \emph {et~al.}(2009)\citenamefont {Baiesi},
  \citenamefont {Maes},\ and\ \citenamefont
  {Wynants}}]{baiesi2009fluctuations}%
  \BibitemOpen
  \bibfield  {author} {\bibinfo {author} {\bibfnamefont {M.}~\bibnamefont
  {Baiesi}}, \bibinfo {author} {\bibfnamefont {C.}~\bibnamefont {Maes}}, \ and\
  \bibinfo {author} {\bibfnamefont {B.}~\bibnamefont {Wynants}},\ }\href@noop
  {} {\bibfield  {journal} {\bibinfo  {journal} {Physical Review Letters}\
  }\textbf {\bibinfo {volume} {103}},\ \bibinfo {pages} {010602} (\bibinfo
  {year} {2009})}\BibitemShut {NoStop}%
\bibitem [{\citenamefont {Maes}(2020{\natexlab{a}})}]{maes}%
  \BibitemOpen
  \bibfield  {author} {\bibinfo {author} {\bibfnamefont {C.}~\bibnamefont
  {Maes}},\ }\href@noop {} {\bibfield  {journal} {\bibinfo  {journal} {Front.
  Phys.}\ }\textbf {\bibinfo {volume} {8}},\ \bibinfo {pages} {00229} (\bibinfo
  {year} {2020}{\natexlab{a}})}\BibitemShut {NoStop}%
\bibitem [{\citenamefont {Lippiello}\ \emph {et~al.}(2008)\citenamefont
  {Lippiello}, \citenamefont {Corberi}, \citenamefont {Sarracino},\ and\
  \citenamefont {Zannetti}}]{PhysRevE.78.041120}%
  \BibitemOpen
  \bibfield  {author} {\bibinfo {author} {\bibfnamefont {E.}~\bibnamefont
  {Lippiello}}, \bibinfo {author} {\bibfnamefont {F.}~\bibnamefont {Corberi}},
  \bibinfo {author} {\bibfnamefont {A.}~\bibnamefont {Sarracino}}, \ and\
  \bibinfo {author} {\bibfnamefont {M.}~\bibnamefont {Zannetti}},\ }\href@noop
  {} {\bibfield  {journal} {\bibinfo  {journal} {Physical Review E}\ }\textbf
  {\bibinfo {volume} {78}},\ \bibinfo {pages} {041120} (\bibinfo {year}
  {2008})}\BibitemShut {NoStop}%
\bibitem [{\citenamefont {Marchetti}\ \emph {et~al.}(2013)\citenamefont
  {Marchetti}, \citenamefont {Joanny}, \citenamefont {Ramaswamy}, \citenamefont
  {Liverpool}, \citenamefont {Prost}, \citenamefont {Rao},\ and\ \citenamefont
  {Simha}}]{marchetti2013hydrodynamics}%
  \BibitemOpen
  \bibfield  {author} {\bibinfo {author} {\bibfnamefont {M.}~\bibnamefont
  {Marchetti}}, \bibinfo {author} {\bibfnamefont {J.}~\bibnamefont {Joanny}},
  \bibinfo {author} {\bibfnamefont {S.}~\bibnamefont {Ramaswamy}}, \bibinfo
  {author} {\bibfnamefont {T.}~\bibnamefont {Liverpool}}, \bibinfo {author}
  {\bibfnamefont {J.}~\bibnamefont {Prost}}, \bibinfo {author} {\bibfnamefont
  {M.}~\bibnamefont {Rao}}, \ and\ \bibinfo {author} {\bibfnamefont {R.~A.}\
  \bibnamefont {Simha}},\ }\href@noop {} {\bibfield  {journal} {\bibinfo
  {journal} {Reviews of Modern Physics}\ }\textbf {\bibinfo {volume} {85}},\
  \bibinfo {pages} {1143} (\bibinfo {year} {2013})}\BibitemShut {NoStop}%
\bibitem [{\citenamefont {Bechinger}\ \emph {et~al.}(2016)\citenamefont
  {Bechinger}, \citenamefont {Di~Leonardo}, \citenamefont {L{\"o}wen},
  \citenamefont {Reichhardt}, \citenamefont {Volpe},\ and\ \citenamefont
  {Volpe}}]{bechinger2016active}%
  \BibitemOpen
  \bibfield  {author} {\bibinfo {author} {\bibfnamefont {C.}~\bibnamefont
  {Bechinger}}, \bibinfo {author} {\bibfnamefont {R.}~\bibnamefont
  {Di~Leonardo}}, \bibinfo {author} {\bibfnamefont {H.}~\bibnamefont
  {L{\"o}wen}}, \bibinfo {author} {\bibfnamefont {C.}~\bibnamefont
  {Reichhardt}}, \bibinfo {author} {\bibfnamefont {G.}~\bibnamefont {Volpe}}, \
  and\ \bibinfo {author} {\bibfnamefont {G.}~\bibnamefont {Volpe}},\
  }\href@noop {} {\bibfield  {journal} {\bibinfo  {journal} {Reviews of Modern
  Physics}\ }\textbf {\bibinfo {volume} {88}},\ \bibinfo {pages} {045006}
  (\bibinfo {year} {2016})}\BibitemShut {NoStop}%
\bibitem [{\citenamefont {Elgeti}\ \emph {et~al.}(2015)\citenamefont {Elgeti},
  \citenamefont {Winkler},\ and\ \citenamefont {Gompper}}]{elgeti2015physics}%
  \BibitemOpen
  \bibfield  {author} {\bibinfo {author} {\bibfnamefont {J.}~\bibnamefont
  {Elgeti}}, \bibinfo {author} {\bibfnamefont {R.~G.}\ \bibnamefont {Winkler}},
  \ and\ \bibinfo {author} {\bibfnamefont {G.}~\bibnamefont {Gompper}},\
  }\href@noop {} {\bibfield  {journal} {\bibinfo  {journal} {Reports on
  progress in physics}\ }\textbf {\bibinfo {volume} {78}},\ \bibinfo {pages}
  {056601} (\bibinfo {year} {2015})}\BibitemShut {NoStop}%
\bibitem [{\citenamefont {Gompper}\ \emph {et~al.}(2020)\citenamefont
  {Gompper}, \citenamefont {Winkler}, \citenamefont {Speck}, \citenamefont
  {Solon}, \citenamefont {Nardini}, \citenamefont {Peruani}, \citenamefont
  {L{\"o}wen}, \citenamefont {Golestanian}, \citenamefont {Kaupp},
  \citenamefont {Alvarez} \emph {et~al.}}]{gompper20202020}%
  \BibitemOpen
  \bibfield  {author} {\bibinfo {author} {\bibfnamefont {G.}~\bibnamefont
  {Gompper}}, \bibinfo {author} {\bibfnamefont {R.~G.}\ \bibnamefont
  {Winkler}}, \bibinfo {author} {\bibfnamefont {T.}~\bibnamefont {Speck}},
  \bibinfo {author} {\bibfnamefont {A.}~\bibnamefont {Solon}}, \bibinfo
  {author} {\bibfnamefont {C.}~\bibnamefont {Nardini}}, \bibinfo {author}
  {\bibfnamefont {F.}~\bibnamefont {Peruani}}, \bibinfo {author} {\bibfnamefont
  {H.}~\bibnamefont {L{\"o}wen}}, \bibinfo {author} {\bibfnamefont
  {R.}~\bibnamefont {Golestanian}}, \bibinfo {author} {\bibfnamefont {U.~B.}\
  \bibnamefont {Kaupp}}, \bibinfo {author} {\bibfnamefont {L.}~\bibnamefont
  {Alvarez}},  \emph {et~al.},\ }\href@noop {} {\bibfield  {journal} {\bibinfo
  {journal} {Journal of Physics: Condensed Matter}\ }\textbf {\bibinfo {volume}
  {32}},\ \bibinfo {pages} {193001} (\bibinfo {year} {2020})}\BibitemShut
  {NoStop}%
\bibitem [{\citenamefont {Shaebani}\ \emph {et~al.}(2020)\citenamefont
  {Shaebani}, \citenamefont {Wysocki}, \citenamefont {Winkler}, \citenamefont
  {Gompper},\ and\ \citenamefont {Rieger}}]{shaebani2020computational}%
  \BibitemOpen
  \bibfield  {author} {\bibinfo {author} {\bibfnamefont {M.~R.}\ \bibnamefont
  {Shaebani}}, \bibinfo {author} {\bibfnamefont {A.}~\bibnamefont {Wysocki}},
  \bibinfo {author} {\bibfnamefont {R.~G.}\ \bibnamefont {Winkler}}, \bibinfo
  {author} {\bibfnamefont {G.}~\bibnamefont {Gompper}}, \ and\ \bibinfo
  {author} {\bibfnamefont {H.}~\bibnamefont {Rieger}},\ }\href@noop {}
  {\bibfield  {journal} {\bibinfo  {journal} {Nature Reviews Physics}\ ,\
  \bibinfo {pages} {1}} (\bibinfo {year} {2020})}\BibitemShut {NoStop}%
\bibitem [{\citenamefont {Fodor}\ and\ \citenamefont
  {Marchetti}(2018)}]{fodor2018statistical}%
  \BibitemOpen
  \bibfield  {author} {\bibinfo {author} {\bibfnamefont {{\'E}.}~\bibnamefont
  {Fodor}}\ and\ \bibinfo {author} {\bibfnamefont {M.~C.}\ \bibnamefont
  {Marchetti}},\ }\href@noop {} {\bibfield  {journal} {\bibinfo  {journal}
  {Physica A: Statistical Mechanics and its Applications}\ }\textbf {\bibinfo
  {volume} {504}},\ \bibinfo {pages} {106} (\bibinfo {year}
  {2018})}\BibitemShut {NoStop}%
\bibitem [{\citenamefont {Caprini}\ \emph
  {et~al.}(2018{\natexlab{a}})\citenamefont {Caprini}, \citenamefont
  {Marconi},\ and\ \citenamefont {Vulpiani}}]{caprini2018linear}%
  \BibitemOpen
  \bibfield  {author} {\bibinfo {author} {\bibfnamefont {L.}~\bibnamefont
  {Caprini}}, \bibinfo {author} {\bibfnamefont {U.~M.~B.}\ \bibnamefont
  {Marconi}}, \ and\ \bibinfo {author} {\bibfnamefont {A.}~\bibnamefont
  {Vulpiani}},\ }\href@noop {} {\bibfield  {journal} {\bibinfo  {journal}
  {Journal of Statistical Mechanics: Theory and Experiment}\ }\textbf {\bibinfo
  {volume} {2018}},\ \bibinfo {pages} {033203} (\bibinfo {year}
  {2018}{\natexlab{a}})}\BibitemShut {NoStop}%
\bibitem [{\citenamefont {Sarracino}\ and\ \citenamefont
  {Vulpiani}(2019)}]{sarracino2019fluctuation}%
  \BibitemOpen
  \bibfield  {author} {\bibinfo {author} {\bibfnamefont {A.}~\bibnamefont
  {Sarracino}}\ and\ \bibinfo {author} {\bibfnamefont {A.}~\bibnamefont
  {Vulpiani}},\ }\href@noop {} {\bibfield  {journal} {\bibinfo  {journal}
  {Chaos: An Interdisciplinary Journal of Nonlinear Science}\ }\textbf
  {\bibinfo {volume} {29}},\ \bibinfo {pages} {083132} (\bibinfo {year}
  {2019})}\BibitemShut {NoStop}%
\bibitem [{\citenamefont {Fodor}\ \emph {et~al.}(2016)\citenamefont {Fodor},
  \citenamefont {Nardini}, \citenamefont {Cates}, \citenamefont {Tailleur},
  \citenamefont {Visco},\ and\ \citenamefont {van Wijland}}]{fodor2016far}%
  \BibitemOpen
  \bibfield  {author} {\bibinfo {author} {\bibfnamefont {{\'E}.}~\bibnamefont
  {Fodor}}, \bibinfo {author} {\bibfnamefont {C.}~\bibnamefont {Nardini}},
  \bibinfo {author} {\bibfnamefont {M.~E.}\ \bibnamefont {Cates}}, \bibinfo
  {author} {\bibfnamefont {J.}~\bibnamefont {Tailleur}}, \bibinfo {author}
  {\bibfnamefont {P.}~\bibnamefont {Visco}}, \ and\ \bibinfo {author}
  {\bibfnamefont {F.}~\bibnamefont {van Wijland}},\ }\href@noop {} {\bibfield
  {journal} {\bibinfo  {journal} {Physical Review Letters}\ }\textbf {\bibinfo
  {volume} {117}},\ \bibinfo {pages} {038103} (\bibinfo {year}
  {2016})}\BibitemShut {NoStop}%
\bibitem [{\citenamefont {Szamel}(2017)}]{szamel2017evaluating}%
  \BibitemOpen
  \bibfield  {author} {\bibinfo {author} {\bibfnamefont {G.}~\bibnamefont
  {Szamel}},\ }\href@noop {} {\bibfield  {journal} {\bibinfo  {journal} {EPL
  (Europhysics Letters)}\ }\textbf {\bibinfo {volume} {117}},\ \bibinfo {pages}
  {50010} (\bibinfo {year} {2017})}\BibitemShut {NoStop}%
\bibitem [{\citenamefont {Berthier}\ and\ \citenamefont
  {Kurchan}(2013)}]{berthier2013non}%
  \BibitemOpen
  \bibfield  {author} {\bibinfo {author} {\bibfnamefont {L.}~\bibnamefont
  {Berthier}}\ and\ \bibinfo {author} {\bibfnamefont {J.}~\bibnamefont
  {Kurchan}},\ }\href@noop {} {\bibfield  {journal} {\bibinfo  {journal}
  {Nature Physics}\ }\textbf {\bibinfo {volume} {9}},\ \bibinfo {pages} {310}
  (\bibinfo {year} {2013})}\BibitemShut {NoStop}%
\bibitem [{\citenamefont {Levis}\ and\ \citenamefont
  {Berthier}(2015)}]{levis2015single}%
  \BibitemOpen
  \bibfield  {author} {\bibinfo {author} {\bibfnamefont {D.}~\bibnamefont
  {Levis}}\ and\ \bibinfo {author} {\bibfnamefont {L.}~\bibnamefont
  {Berthier}},\ }\href@noop {} {\bibfield  {journal} {\bibinfo  {journal} {EPL
  (Europhysics Letters)}\ }\textbf {\bibinfo {volume} {111}},\ \bibinfo {pages}
  {60006} (\bibinfo {year} {2015})}\BibitemShut {NoStop}%
\bibitem [{\citenamefont {Nandi}\ and\ \citenamefont
  {Gov}(2018)}]{nandi2018effective}%
  \BibitemOpen
  \bibfield  {author} {\bibinfo {author} {\bibfnamefont {S.~K.}\ \bibnamefont
  {Nandi}}\ and\ \bibinfo {author} {\bibfnamefont {N.}~\bibnamefont {Gov}},\
  }\href@noop {} {\bibfield  {journal} {\bibinfo  {journal} {The European
  Physical Journal E}\ }\textbf {\bibinfo {volume} {41}},\ \bibinfo {pages}
  {117} (\bibinfo {year} {2018})}\BibitemShut {NoStop}%
\bibitem [{\citenamefont {Cugliandolo}\ \emph {et~al.}(2019)\citenamefont
  {Cugliandolo}, \citenamefont {Gonnella},\ and\ \citenamefont
  {Petrelli}}]{cugliandolo2019effective}%
  \BibitemOpen
  \bibfield  {author} {\bibinfo {author} {\bibfnamefont {L.~F.}\ \bibnamefont
  {Cugliandolo}}, \bibinfo {author} {\bibfnamefont {G.}~\bibnamefont
  {Gonnella}}, \ and\ \bibinfo {author} {\bibfnamefont {I.}~\bibnamefont
  {Petrelli}},\ }\href@noop {} {\bibfield  {journal} {\bibinfo  {journal}
  {Fluctuation and Noise Letters}\ }\textbf {\bibinfo {volume} {18}},\ \bibinfo
  {pages} {1940008} (\bibinfo {year} {2019})}\BibitemShut {NoStop}%
\bibitem [{\citenamefont {Preisler}\ and\ \citenamefont
  {Dijkstra}(2016)}]{preisler2016configurational}%
  \BibitemOpen
  \bibfield  {author} {\bibinfo {author} {\bibfnamefont {Z.}~\bibnamefont
  {Preisler}}\ and\ \bibinfo {author} {\bibfnamefont {M.}~\bibnamefont
  {Dijkstra}},\ }\href@noop {} {\bibfield  {journal} {\bibinfo  {journal} {Soft
  Matter}\ }\textbf {\bibinfo {volume} {12}},\ \bibinfo {pages} {6043}
  (\bibinfo {year} {2016})}\BibitemShut {NoStop}%
\bibitem [{\citenamefont {Petrelli}\ \emph {et~al.}(2020)\citenamefont
  {Petrelli}, \citenamefont {Cugliandolo}, \citenamefont {Gonnella},\ and\
  \citenamefont {Suma}}]{petrelli2020effective}%
  \BibitemOpen
  \bibfield  {author} {\bibinfo {author} {\bibfnamefont {I.}~\bibnamefont
  {Petrelli}}, \bibinfo {author} {\bibfnamefont {L.~F.}\ \bibnamefont
  {Cugliandolo}}, \bibinfo {author} {\bibfnamefont {G.}~\bibnamefont
  {Gonnella}}, \ and\ \bibinfo {author} {\bibfnamefont {A.}~\bibnamefont
  {Suma}},\ }\href {\doibase 10.1103/PhysRevE.102.012609} {\bibfield  {journal}
  {\bibinfo  {journal} {Physical Review E}\ }\textbf {\bibinfo {volume}
  {102}},\ \bibinfo {pages} {012609} (\bibinfo {year} {2020})}\BibitemShut
  {NoStop}%
\bibitem [{\citenamefont {Villamaina}\ \emph {et~al.}(2009)\citenamefont
  {Villamaina}, \citenamefont {Baldassarri}, \citenamefont {Puglisi},\ and\
  \citenamefont {Vulpiani}}]{VBPV09}%
  \BibitemOpen
  \bibfield  {author} {\bibinfo {author} {\bibfnamefont {D.}~\bibnamefont
  {Villamaina}}, \bibinfo {author} {\bibfnamefont {A.}~\bibnamefont
  {Baldassarri}}, \bibinfo {author} {\bibfnamefont {A.}~\bibnamefont
  {Puglisi}}, \ and\ \bibinfo {author} {\bibfnamefont {A.}~\bibnamefont
  {Vulpiani}},\ }\href@noop {} {\bibfield  {journal} {\bibinfo  {journal} {J.
  Stat. Mech.}\ ,\ \bibinfo {pages} {P07024}} (\bibinfo {year}
  {2009})}\BibitemShut {NoStop}%
\bibitem [{\citenamefont {Dal~Cengio}\ \emph {et~al.}(2019)\citenamefont
  {Dal~Cengio}, \citenamefont {Levis},\ and\ \citenamefont
  {Pagonabarraga}}]{dal2019linear}%
  \BibitemOpen
  \bibfield  {author} {\bibinfo {author} {\bibfnamefont {S.}~\bibnamefont
  {Dal~Cengio}}, \bibinfo {author} {\bibfnamefont {D.}~\bibnamefont {Levis}}, \
  and\ \bibinfo {author} {\bibfnamefont {I.}~\bibnamefont {Pagonabarraga}},\
  }\href@noop {} {\bibfield  {journal} {\bibinfo  {journal} {Physical Review
  Letters}\ }\textbf {\bibinfo {volume} {123}},\ \bibinfo {pages} {238003}
  (\bibinfo {year} {2019})}\BibitemShut {NoStop}%
\bibitem [{\citenamefont {Dal~Cengio}\ \emph {et~al.}(2020)\citenamefont
  {Dal~Cengio}, \citenamefont {Levis},\ and\ \citenamefont
  {Pagonabarraga}}]{cengio2020fluctuation}%
  \BibitemOpen
  \bibfield  {author} {\bibinfo {author} {\bibfnamefont {S.}~\bibnamefont
  {Dal~Cengio}}, \bibinfo {author} {\bibfnamefont {D.}~\bibnamefont {Levis}}, \
  and\ \bibinfo {author} {\bibfnamefont {I.}~\bibnamefont {Pagonabarraga}},\
  }\href@noop {} {\bibfield  {journal} {\bibinfo  {journal} {arXiv preprint
  arXiv:2007.07322}\ } (\bibinfo {year} {2020})}\BibitemShut {NoStop}%
\bibitem [{\citenamefont {Burkholdera}\ and\ \citenamefont
  {Brady}(2019)}]{brady}%
  \BibitemOpen
  \bibfield  {author} {\bibinfo {author} {\bibfnamefont {E.~W.}\ \bibnamefont
  {Burkholdera}}\ and\ \bibinfo {author} {\bibfnamefont {J.~F.}\ \bibnamefont
  {Brady}},\ }\href@noop {} {\bibfield  {journal} {\bibinfo  {journal} {The
  Journal of Chemical Physics}\ }\textbf {\bibinfo {volume} {150}},\ \bibinfo
  {pages} {184901} (\bibinfo {year} {2019})}\BibitemShut {NoStop}%
\bibitem [{\citenamefont {Maes}(2020{\natexlab{b}})}]{maes2020fluctuating}%
  \BibitemOpen
  \bibfield  {author} {\bibinfo {author} {\bibfnamefont {C.}~\bibnamefont
  {Maes}},\ }\href@noop {} {\bibfield  {journal} {\bibinfo  {journal} {Physical
  Review Letters}\ }\textbf {\bibinfo {volume} {125}} (\bibinfo {year}
  {2020}{\natexlab{b}})}\BibitemShut {NoStop}%
\bibitem [{\citenamefont {Berthier}\ \emph {et~al.}(2017)\citenamefont
  {Berthier}, \citenamefont {Flenner},\ and\ \citenamefont
  {Szamel}}]{berthier2017active}%
  \BibitemOpen
  \bibfield  {author} {\bibinfo {author} {\bibfnamefont {L.}~\bibnamefont
  {Berthier}}, \bibinfo {author} {\bibfnamefont {E.}~\bibnamefont {Flenner}}, \
  and\ \bibinfo {author} {\bibfnamefont {G.}~\bibnamefont {Szamel}},\
  }\href@noop {} {\bibfield  {journal} {\bibinfo  {journal} {New J. Phys.}\
  }\textbf {\bibinfo {volume} {19}},\ \bibinfo {pages} {125006} (\bibinfo
  {year} {2017})}\BibitemShut {NoStop}%
\bibitem [{\citenamefont {Mandal}\ \emph {et~al.}(2017)\citenamefont {Mandal},
  \citenamefont {Klymko},\ and\ \citenamefont {DeWeese}}]{mandal2017entropy}%
  \BibitemOpen
  \bibfield  {author} {\bibinfo {author} {\bibfnamefont {D.}~\bibnamefont
  {Mandal}}, \bibinfo {author} {\bibfnamefont {K.}~\bibnamefont {Klymko}}, \
  and\ \bibinfo {author} {\bibfnamefont {M.~R.}\ \bibnamefont {DeWeese}},\
  }\href@noop {} {\bibfield  {journal} {\bibinfo  {journal} {Physical Review
  Letters}\ }\textbf {\bibinfo {volume} {119}},\ \bibinfo {pages} {258001}
  (\bibinfo {year} {2017})}\BibitemShut {NoStop}%
\bibitem [{\citenamefont {Caprini}\ and\ \citenamefont
  {Marconi}(2018)}]{caprini2018active}%
  \BibitemOpen
  \bibfield  {author} {\bibinfo {author} {\bibfnamefont {L.}~\bibnamefont
  {Caprini}}\ and\ \bibinfo {author} {\bibfnamefont {U.~M.~B.}\ \bibnamefont
  {Marconi}},\ }\href@noop {} {\bibfield  {journal} {\bibinfo  {journal} {Soft
  Matter}\ }\textbf {\bibinfo {volume} {14}},\ \bibinfo {pages} {9044}
  (\bibinfo {year} {2018})}\BibitemShut {NoStop}%
\bibitem [{\citenamefont {Wittmann}\ \emph {et~al.}(2018)\citenamefont
  {Wittmann}, \citenamefont {Brader}, \citenamefont {Sharma},\ and\
  \citenamefont {Marconi}}]{wittmann2018effective}%
  \BibitemOpen
  \bibfield  {author} {\bibinfo {author} {\bibfnamefont {R.}~\bibnamefont
  {Wittmann}}, \bibinfo {author} {\bibfnamefont {J.~M.}\ \bibnamefont
  {Brader}}, \bibinfo {author} {\bibfnamefont {A.}~\bibnamefont {Sharma}}, \
  and\ \bibinfo {author} {\bibfnamefont {U.~M.~B.}\ \bibnamefont {Marconi}},\
  }\href@noop {} {\bibfield  {journal} {\bibinfo  {journal} {Physical Review
  E}\ }\textbf {\bibinfo {volume} {97}},\ \bibinfo {pages} {012601} (\bibinfo
  {year} {2018})}\BibitemShut {NoStop}%
\bibitem [{\citenamefont {Bonilla}(2019)}]{bonilla2019active}%
  \BibitemOpen
  \bibfield  {author} {\bibinfo {author} {\bibfnamefont {L.~L.}\ \bibnamefont
  {Bonilla}},\ }\href@noop {} {\bibfield  {journal} {\bibinfo  {journal}
  {Physical Review E}\ }\textbf {\bibinfo {volume} {100}},\ \bibinfo {pages}
  {022601} (\bibinfo {year} {2019})}\BibitemShut {NoStop}%
\bibitem [{\citenamefont {Dabelow}\ \emph {et~al.}(2019)\citenamefont
  {Dabelow}, \citenamefont {Bo},\ and\ \citenamefont
  {Eichhorn}}]{dabelow2019irreversibility}%
  \BibitemOpen
  \bibfield  {author} {\bibinfo {author} {\bibfnamefont {L.}~\bibnamefont
  {Dabelow}}, \bibinfo {author} {\bibfnamefont {S.}~\bibnamefont {Bo}}, \ and\
  \bibinfo {author} {\bibfnamefont {R.}~\bibnamefont {Eichhorn}},\ }\href@noop
  {} {\bibfield  {journal} {\bibinfo  {journal} {Physical Review X}\ }\textbf
  {\bibinfo {volume} {9}},\ \bibinfo {pages} {021009} (\bibinfo {year}
  {2019})}\BibitemShut {NoStop}%
\bibitem [{\citenamefont {Martin}\ \emph {et~al.}(2020)\citenamefont {Martin},
  \citenamefont {O'Byrne}, \citenamefont {Cates}, \citenamefont {Fodor},
  \citenamefont {Nardini}, \citenamefont {Tailleur},\ and\ \citenamefont {van
  Wijland}}]{martin2020statistical}%
  \BibitemOpen
  \bibfield  {author} {\bibinfo {author} {\bibfnamefont {D.}~\bibnamefont
  {Martin}}, \bibinfo {author} {\bibfnamefont {J.}~\bibnamefont {O'Byrne}},
  \bibinfo {author} {\bibfnamefont {M.~E.}\ \bibnamefont {Cates}}, \bibinfo
  {author} {\bibfnamefont {{\'E}.}~\bibnamefont {Fodor}}, \bibinfo {author}
  {\bibfnamefont {C.}~\bibnamefont {Nardini}}, \bibinfo {author} {\bibfnamefont
  {J.}~\bibnamefont {Tailleur}}, \ and\ \bibinfo {author} {\bibfnamefont
  {F.}~\bibnamefont {van Wijland}},\ }\href@noop {} {\bibfield  {journal}
  {\bibinfo  {journal} {arXiv preprint arXiv:2008.12972}\ } (\bibinfo {year}
  {2020})}\BibitemShut {NoStop}%
\bibitem [{\citenamefont {Woillez}\ \emph
  {et~al.}(2020{\natexlab{a}})\citenamefont {Woillez}, \citenamefont {Kafri},\
  and\ \citenamefont {Gov}}]{woillez2020active}%
  \BibitemOpen
  \bibfield  {author} {\bibinfo {author} {\bibfnamefont {E.}~\bibnamefont
  {Woillez}}, \bibinfo {author} {\bibfnamefont {Y.}~\bibnamefont {Kafri}}, \
  and\ \bibinfo {author} {\bibfnamefont {N.~S.}\ \bibnamefont {Gov}},\
  }\href@noop {} {\bibfield  {journal} {\bibinfo  {journal} {Physical Review
  Letters}\ }\textbf {\bibinfo {volume} {124}},\ \bibinfo {pages} {118002}
  (\bibinfo {year} {2020}{\natexlab{a}})}\BibitemShut {NoStop}%
\bibitem [{\citenamefont {Wu}\ and\ \citenamefont
  {Libchaber}(2000)}]{wu2000particle}%
  \BibitemOpen
  \bibfield  {author} {\bibinfo {author} {\bibfnamefont {X.-L.}\ \bibnamefont
  {Wu}}\ and\ \bibinfo {author} {\bibfnamefont {A.}~\bibnamefont {Libchaber}},\
  }\href@noop {} {\bibfield  {journal} {\bibinfo  {journal} {Physical Review
  Letters}\ }\textbf {\bibinfo {volume} {84}},\ \bibinfo {pages} {3017}
  (\bibinfo {year} {2000})}\BibitemShut {NoStop}%
\bibitem [{\citenamefont {Maggi}\ \emph {et~al.}(2014)\citenamefont {Maggi},
  \citenamefont {Paoluzzi}, \citenamefont {Pellicciotta}, \citenamefont
  {Lepore}, \citenamefont {Angelani},\ and\ \citenamefont
  {Di~Leonardo}}]{maggi2014generalized}%
  \BibitemOpen
  \bibfield  {author} {\bibinfo {author} {\bibfnamefont {C.}~\bibnamefont
  {Maggi}}, \bibinfo {author} {\bibfnamefont {M.}~\bibnamefont {Paoluzzi}},
  \bibinfo {author} {\bibfnamefont {N.}~\bibnamefont {Pellicciotta}}, \bibinfo
  {author} {\bibfnamefont {A.}~\bibnamefont {Lepore}}, \bibinfo {author}
  {\bibfnamefont {L.}~\bibnamefont {Angelani}}, \ and\ \bibinfo {author}
  {\bibfnamefont {R.}~\bibnamefont {Di~Leonardo}},\ }\href@noop {} {\bibfield
  {journal} {\bibinfo  {journal} {Physical Review Letters}\ }\textbf {\bibinfo
  {volume} {113}},\ \bibinfo {pages} {238303} (\bibinfo {year}
  {2014})}\BibitemShut {NoStop}%
\bibitem [{\citenamefont {Maggi}\ \emph {et~al.}(2017)\citenamefont {Maggi},
  \citenamefont {Paoluzzi}, \citenamefont {Angelani},\ and\ \citenamefont
  {Di~Leonardo}}]{maggi2017memory}%
  \BibitemOpen
  \bibfield  {author} {\bibinfo {author} {\bibfnamefont {C.}~\bibnamefont
  {Maggi}}, \bibinfo {author} {\bibfnamefont {M.}~\bibnamefont {Paoluzzi}},
  \bibinfo {author} {\bibfnamefont {L.}~\bibnamefont {Angelani}}, \ and\
  \bibinfo {author} {\bibfnamefont {R.}~\bibnamefont {Di~Leonardo}},\
  }\href@noop {} {\bibfield  {journal} {\bibinfo  {journal} {Scientific
  Reports}\ }\textbf {\bibinfo {volume} {7}},\ \bibinfo {pages} {1} (\bibinfo
  {year} {2017})}\BibitemShut {NoStop}%
\bibitem [{\citenamefont {Caprini}\ \emph
  {et~al.}(2019{\natexlab{a}})\citenamefont {Caprini}, \citenamefont
  {Hern{\'a}ndez-Garc{\'\i}a}, \citenamefont {L{\'o}pez},\ and\ \citenamefont
  {Marconi}}]{caprini2019comparative}%
  \BibitemOpen
  \bibfield  {author} {\bibinfo {author} {\bibfnamefont {L.}~\bibnamefont
  {Caprini}}, \bibinfo {author} {\bibfnamefont {E.}~\bibnamefont
  {Hern{\'a}ndez-Garc{\'\i}a}}, \bibinfo {author} {\bibfnamefont
  {C.}~\bibnamefont {L{\'o}pez}}, \ and\ \bibinfo {author} {\bibfnamefont
  {U.~M.~B.}\ \bibnamefont {Marconi}},\ }\href@noop {} {\bibfield  {journal}
  {\bibinfo  {journal} {Scientific Reports}\ }\textbf {\bibinfo {volume} {9}},\
  \bibinfo {pages} {1} (\bibinfo {year} {2019}{\natexlab{a}})}\BibitemShut
  {NoStop}%
\bibitem [{\citenamefont {Das}\ \emph {et~al.}(2018)\citenamefont {Das},
  \citenamefont {Gompper},\ and\ \citenamefont {Winkler}}]{das2018confined}%
  \BibitemOpen
  \bibfield  {author} {\bibinfo {author} {\bibfnamefont {S.}~\bibnamefont
  {Das}}, \bibinfo {author} {\bibfnamefont {G.}~\bibnamefont {Gompper}}, \ and\
  \bibinfo {author} {\bibfnamefont {R.~G.}\ \bibnamefont {Winkler}},\
  }\href@noop {} {\bibfield  {journal} {\bibinfo  {journal} {New Journal of
  Physics}\ }\textbf {\bibinfo {volume} {20}},\ \bibinfo {pages} {015001}
  (\bibinfo {year} {2018})}\BibitemShut {NoStop}%
\bibitem [{\citenamefont {Caprini}\ and\ \citenamefont
  {Marconi}(2019)}]{caprini2019active}%
  \BibitemOpen
  \bibfield  {author} {\bibinfo {author} {\bibfnamefont {L.}~\bibnamefont
  {Caprini}}\ and\ \bibinfo {author} {\bibfnamefont {U.~M.~B.}\ \bibnamefont
  {Marconi}},\ }\href@noop {} {\bibfield  {journal} {\bibinfo  {journal} {Soft
  Matter}\ }\textbf {\bibinfo {volume} {15}},\ \bibinfo {pages} {2627}
  (\bibinfo {year} {2019})}\BibitemShut {NoStop}%
\bibitem [{\citenamefont {Farage}\ \emph {et~al.}(2015)\citenamefont {Farage},
  \citenamefont {Krinninger},\ and\ \citenamefont
  {Brader}}]{farage2015effective}%
  \BibitemOpen
  \bibfield  {author} {\bibinfo {author} {\bibfnamefont {T.~F.}\ \bibnamefont
  {Farage}}, \bibinfo {author} {\bibfnamefont {P.}~\bibnamefont {Krinninger}},
  \ and\ \bibinfo {author} {\bibfnamefont {J.~M.}\ \bibnamefont {Brader}},\
  }\href@noop {} {\bibfield  {journal} {\bibinfo  {journal} {Physical Review
  E}\ }\textbf {\bibinfo {volume} {91}},\ \bibinfo {pages} {042310} (\bibinfo
  {year} {2015})}\BibitemShut {NoStop}%
\bibitem [{\citenamefont {Maggi}\ \emph {et~al.}(2020)\citenamefont {Maggi},
  \citenamefont {Paoluzzi}, \citenamefont {Crisanti}, \citenamefont
  {Zaccarelli},\ and\ \citenamefont {Gnan}}]{maggi2020universality}%
  \BibitemOpen
  \bibfield  {author} {\bibinfo {author} {\bibfnamefont {C.}~\bibnamefont
  {Maggi}}, \bibinfo {author} {\bibfnamefont {M.}~\bibnamefont {Paoluzzi}},
  \bibinfo {author} {\bibfnamefont {A.}~\bibnamefont {Crisanti}}, \bibinfo
  {author} {\bibfnamefont {E.}~\bibnamefont {Zaccarelli}}, \ and\ \bibinfo
  {author} {\bibfnamefont {N.}~\bibnamefont {Gnan}},\ }\href@noop {} {\bibfield
   {journal} {\bibinfo  {journal} {arXiv preprint arXiv:2007.12660}\ }
  (\bibinfo {year} {2020})}\BibitemShut {NoStop}%
\bibitem [{\citenamefont {Caprini}\ \emph {et~al.}(2020)\citenamefont
  {Caprini}, \citenamefont {Marconi}, \citenamefont {Maggi}, \citenamefont
  {Paoluzzi},\ and\ \citenamefont {Puglisi}}]{caprini2020hidden}%
  \BibitemOpen
  \bibfield  {author} {\bibinfo {author} {\bibfnamefont {L.}~\bibnamefont
  {Caprini}}, \bibinfo {author} {\bibfnamefont {U.~M.~B.}\ \bibnamefont
  {Marconi}}, \bibinfo {author} {\bibfnamefont {C.}~\bibnamefont {Maggi}},
  \bibinfo {author} {\bibfnamefont {M.}~\bibnamefont {Paoluzzi}}, \ and\
  \bibinfo {author} {\bibfnamefont {A.}~\bibnamefont {Puglisi}},\ }\href@noop
  {} {\bibfield  {journal} {\bibinfo  {journal} {Physical Review Research}\
  }\textbf {\bibinfo {volume} {2}},\ \bibinfo {pages} {023321} (\bibinfo {year}
  {2020})}\BibitemShut {NoStop}%
\bibitem [{\citenamefont {Caprini}\ and\ \citenamefont
  {Marconi}(2020)}]{caprini2020time}%
  \BibitemOpen
  \bibfield  {author} {\bibinfo {author} {\bibfnamefont {L.}~\bibnamefont
  {Caprini}}\ and\ \bibinfo {author} {\bibfnamefont {U.~M.~B.}\ \bibnamefont
  {Marconi}},\ }\href@noop {} {\bibfield  {journal} {\bibinfo  {journal}
  {Physical Review Research}\ }\textbf {\bibinfo {volume} {2}},\ \bibinfo
  {pages} {033518} (\bibinfo {year} {2020})}\BibitemShut {NoStop}%
\bibitem [{\citenamefont {Puglisi}\ and\ \citenamefont {Marini
  Bettolo~Marconi}(2017)}]{puglisi2017clausius}%
  \BibitemOpen
  \bibfield  {author} {\bibinfo {author} {\bibfnamefont {A.}~\bibnamefont
  {Puglisi}}\ and\ \bibinfo {author} {\bibfnamefont {U.}~\bibnamefont {Marini
  Bettolo~Marconi}},\ }\href@noop {} {\bibfield  {journal} {\bibinfo  {journal}
  {Entropy}\ }\textbf {\bibinfo {volume} {19}},\ \bibinfo {pages} {356}
  (\bibinfo {year} {2017})}\BibitemShut {NoStop}%
\bibitem [{\citenamefont {Caprini}\ \emph
  {et~al.}(2019{\natexlab{b}})\citenamefont {Caprini}, \citenamefont {Marconi},
  \citenamefont {Puglisi},\ and\ \citenamefont
  {Vulpiani}}]{caprini2019entropy}%
  \BibitemOpen
  \bibfield  {author} {\bibinfo {author} {\bibfnamefont {L.}~\bibnamefont
  {Caprini}}, \bibinfo {author} {\bibfnamefont {U.~M.~B.}\ \bibnamefont
  {Marconi}}, \bibinfo {author} {\bibfnamefont {A.}~\bibnamefont {Puglisi}}, \
  and\ \bibinfo {author} {\bibfnamefont {A.}~\bibnamefont {Vulpiani}},\
  }\href@noop {} {\bibfield  {journal} {\bibinfo  {journal} {Journal of
  Statistical Mechanics: Theory and Experiment}\ }\textbf {\bibinfo {volume}
  {2019}},\ \bibinfo {pages} {053203} (\bibinfo {year}
  {2019}{\natexlab{b}})}\BibitemShut {NoStop}%
\bibitem [{\citenamefont {Dabelow}\ and\ \citenamefont
  {Eichhorn}(2020)}]{dabelow2020irreversibility}%
  \BibitemOpen
  \bibfield  {author} {\bibinfo {author} {\bibfnamefont {L.}~\bibnamefont
  {Dabelow}}\ and\ \bibinfo {author} {\bibfnamefont {R.}~\bibnamefont
  {Eichhorn}},\ }\href@noop {} {\bibfield  {journal} {\bibinfo  {journal}
  {arXiv preprint arXiv:2011.02976}\ } (\bibinfo {year} {2020})}\BibitemShut
  {NoStop}%
\bibitem [{\citenamefont {Marconi}\ \emph {et~al.}(2017)\citenamefont
  {Marconi}, \citenamefont {Puglisi},\ and\ \citenamefont
  {Maggi}}]{marconi2017heat}%
  \BibitemOpen
  \bibfield  {author} {\bibinfo {author} {\bibfnamefont {U.~M.~B.}\
  \bibnamefont {Marconi}}, \bibinfo {author} {\bibfnamefont {A.}~\bibnamefont
  {Puglisi}}, \ and\ \bibinfo {author} {\bibfnamefont {C.}~\bibnamefont
  {Maggi}},\ }\href@noop {} {\bibfield  {journal} {\bibinfo  {journal}
  {Scientific Reports}\ }\textbf {\bibinfo {volume} {7}},\ \bibinfo {pages}
  {46496} (\bibinfo {year} {2017})}\BibitemShut {NoStop}%
\bibitem [{\citenamefont {Caprini}\ \emph
  {et~al.}(2018{\natexlab{b}})\citenamefont {Caprini}, \citenamefont {Marconi},
  \citenamefont {Puglisi},\ and\ \citenamefont
  {Vulpiani}}]{caprini2018comment}%
  \BibitemOpen
  \bibfield  {author} {\bibinfo {author} {\bibfnamefont {L.}~\bibnamefont
  {Caprini}}, \bibinfo {author} {\bibfnamefont {U.~M.~B.}\ \bibnamefont
  {Marconi}}, \bibinfo {author} {\bibfnamefont {A.}~\bibnamefont {Puglisi}}, \
  and\ \bibinfo {author} {\bibfnamefont {A.}~\bibnamefont {Vulpiani}},\
  }\href@noop {} {\bibfield  {journal} {\bibinfo  {journal} {Physical Review
  Letters}\ }\textbf {\bibinfo {volume} {121}},\ \bibinfo {pages} {139801}
  (\bibinfo {year} {2018}{\natexlab{b}})}\BibitemShut {NoStop}%
\bibitem [{\citenamefont {Martin}(2020)}]{martin2020aoup}%
  \BibitemOpen
  \bibfield  {author} {\bibinfo {author} {\bibfnamefont {D.}~\bibnamefont
  {Martin}},\ }\href@noop {} {\bibfield  {journal} {\bibinfo  {journal} {arXiv
  preprint arXiv:2009.13476}\ } (\bibinfo {year} {2020})}\BibitemShut {NoStop}%
\bibitem [{\citenamefont {Szamel}(2014)}]{szamel2014self}%
  \BibitemOpen
  \bibfield  {author} {\bibinfo {author} {\bibfnamefont {G.}~\bibnamefont
  {Szamel}},\ }\href@noop {} {\bibfield  {journal} {\bibinfo  {journal}
  {Physical Review E}\ }\textbf {\bibinfo {volume} {90}},\ \bibinfo {pages}
  {012111} (\bibinfo {year} {2014})}\BibitemShut {NoStop}%
\bibitem [{\citenamefont {Caprini}\ \emph
  {et~al.}(2019{\natexlab{c}})\citenamefont {Caprini}, \citenamefont {Marini
  Bettolo~Marconi}, \citenamefont {Puglisi},\ and\ \citenamefont
  {Vulpiani}}]{caprini2019activedoublewell}%
  \BibitemOpen
  \bibfield  {author} {\bibinfo {author} {\bibfnamefont {L.}~\bibnamefont
  {Caprini}}, \bibinfo {author} {\bibfnamefont {U.}~\bibnamefont {Marini
  Bettolo~Marconi}}, \bibinfo {author} {\bibfnamefont {A.}~\bibnamefont
  {Puglisi}}, \ and\ \bibinfo {author} {\bibfnamefont {A.}~\bibnamefont
  {Vulpiani}},\ }\href@noop {} {\bibfield  {journal} {\bibinfo  {journal} {The
  Journal of Chemical Physics}\ }\textbf {\bibinfo {volume} {150}},\ \bibinfo
  {pages} {024902} (\bibinfo {year} {2019}{\natexlab{c}})}\BibitemShut
  {NoStop}%
\bibitem [{\citenamefont {Woillez}\ \emph
  {et~al.}(2020{\natexlab{b}})\citenamefont {Woillez}, \citenamefont {Kafri},\
  and\ \citenamefont {Lecomte}}]{woillez2020nonlocal}%
  \BibitemOpen
  \bibfield  {author} {\bibinfo {author} {\bibfnamefont {E.}~\bibnamefont
  {Woillez}}, \bibinfo {author} {\bibfnamefont {Y.}~\bibnamefont {Kafri}}, \
  and\ \bibinfo {author} {\bibfnamefont {V.}~\bibnamefont {Lecomte}},\
  }\href@noop {} {\bibfield  {journal} {\bibinfo  {journal} {Journal of
  Statistical Mechanics: Theory and Experiment}\ }\textbf {\bibinfo {volume}
  {2020}},\ \bibinfo {pages} {063204} (\bibinfo {year}
  {2020}{\natexlab{b}})}\BibitemShut {NoStop}%
\bibitem [{\citenamefont {Caprini}\ \emph
  {et~al.}(2019{\natexlab{d}})\citenamefont {Caprini}, \citenamefont
  {Marconi},\ and\ \citenamefont {Puglisi}}]{caprini2019activity}%
  \BibitemOpen
  \bibfield  {author} {\bibinfo {author} {\bibfnamefont {L.}~\bibnamefont
  {Caprini}}, \bibinfo {author} {\bibfnamefont {U.~M.~B.}\ \bibnamefont
  {Marconi}}, \ and\ \bibinfo {author} {\bibfnamefont {A.}~\bibnamefont
  {Puglisi}},\ }\href@noop {} {\bibfield  {journal} {\bibinfo  {journal}
  {Scientific Reports}\ }\textbf {\bibinfo {volume} {9}},\ \bibinfo {pages} {1}
  (\bibinfo {year} {2019}{\natexlab{d}})}\BibitemShut {NoStop}%
\bibitem [{\citenamefont {Stenhammar}\ \emph {et~al.}(2014)\citenamefont
  {Stenhammar}, \citenamefont {Marenduzzo}, \citenamefont {Allen},\ and\
  \citenamefont {Cates}}]{stenhammar2014phase}%
  \BibitemOpen
  \bibfield  {author} {\bibinfo {author} {\bibfnamefont {J.}~\bibnamefont
  {Stenhammar}}, \bibinfo {author} {\bibfnamefont {D.}~\bibnamefont
  {Marenduzzo}}, \bibinfo {author} {\bibfnamefont {R.~J.}\ \bibnamefont
  {Allen}}, \ and\ \bibinfo {author} {\bibfnamefont {M.~E.}\ \bibnamefont
  {Cates}},\ }\href@noop {} {\bibfield  {journal} {\bibinfo  {journal} {Soft
  Matter}\ }\textbf {\bibinfo {volume} {10}},\ \bibinfo {pages} {1489}
  (\bibinfo {year} {2014})}\BibitemShut {NoStop}%
\bibitem [{\citenamefont {Fily}(2019)}]{fily2019self}%
  \BibitemOpen
  \bibfield  {author} {\bibinfo {author} {\bibfnamefont {Y.}~\bibnamefont
  {Fily}},\ }\href@noop {} {\bibfield  {journal} {\bibinfo  {journal} {The
  Journal of Chemical Physics}\ }\textbf {\bibinfo {volume} {150}},\ \bibinfo
  {pages} {174906} (\bibinfo {year} {2019})}\BibitemShut {NoStop}%
\bibitem [{\citenamefont {Wio}\ \emph {et~al.}(1989)\citenamefont {Wio},
  \citenamefont {Colet}, \citenamefont {San~Miguel}, \citenamefont {Pesquera},\
  and\ \citenamefont {Rodriguez}}]{wio1989path}%
  \BibitemOpen
  \bibfield  {author} {\bibinfo {author} {\bibfnamefont {H.~S.}\ \bibnamefont
  {Wio}}, \bibinfo {author} {\bibfnamefont {P.}~\bibnamefont {Colet}}, \bibinfo
  {author} {\bibfnamefont {M.}~\bibnamefont {San~Miguel}}, \bibinfo {author}
  {\bibfnamefont {L.}~\bibnamefont {Pesquera}}, \ and\ \bibinfo {author}
  {\bibfnamefont {M.}~\bibnamefont {Rodriguez}},\ }\href@noop {} {\bibfield
  {journal} {\bibinfo  {journal} {Physical Review A}\ }\textbf {\bibinfo
  {volume} {40}},\ \bibinfo {pages} {7312} (\bibinfo {year}
  {1989})}\BibitemShut {NoStop}%
\bibitem [{\citenamefont {Bray}\ \emph {et~al.}(1990)\citenamefont {Bray},
  \citenamefont {McKane},\ and\ \citenamefont {Newman}}]{bray1990path}%
  \BibitemOpen
  \bibfield  {author} {\bibinfo {author} {\bibfnamefont {A.}~\bibnamefont
  {Bray}}, \bibinfo {author} {\bibfnamefont {A.}~\bibnamefont {McKane}}, \ and\
  \bibinfo {author} {\bibfnamefont {T.}~\bibnamefont {Newman}},\ }\href@noop {}
  {\bibfield  {journal} {\bibinfo  {journal} {Physical Review A}\ }\textbf
  {\bibinfo {volume} {41}},\ \bibinfo {pages} {657} (\bibinfo {year}
  {1990})}\BibitemShut {NoStop}%
\bibitem [{\citenamefont {Sharma}\ \emph {et~al.}(2017)\citenamefont {Sharma},
  \citenamefont {Wittmann},\ and\ \citenamefont {Brader}}]{sharma2017escape}%
  \BibitemOpen
  \bibfield  {author} {\bibinfo {author} {\bibfnamefont {A.}~\bibnamefont
  {Sharma}}, \bibinfo {author} {\bibfnamefont {R.}~\bibnamefont {Wittmann}}, \
  and\ \bibinfo {author} {\bibfnamefont {J.~M.}\ \bibnamefont {Brader}},\
  }\href@noop {} {\bibfield  {journal} {\bibinfo  {journal} {Physical Review
  E}\ }\textbf {\bibinfo {volume} {95}},\ \bibinfo {pages} {012115} (\bibinfo
  {year} {2017})}\BibitemShut {NoStop}%
\bibitem [{\citenamefont {Maggi}\ \emph {et~al.}(2015)\citenamefont {Maggi},
  \citenamefont {Marconi}, \citenamefont {Gnan},\ and\ \citenamefont
  {Di~Leonardo}}]{maggi2015multidimensional}%
  \BibitemOpen
  \bibfield  {author} {\bibinfo {author} {\bibfnamefont {C.}~\bibnamefont
  {Maggi}}, \bibinfo {author} {\bibfnamefont {U.~M.~B.}\ \bibnamefont
  {Marconi}}, \bibinfo {author} {\bibfnamefont {N.}~\bibnamefont {Gnan}}, \
  and\ \bibinfo {author} {\bibfnamefont {R.}~\bibnamefont {Di~Leonardo}},\
  }\href@noop {} {\bibfield  {journal} {\bibinfo  {journal} {Scientific
  Reports}\ }\textbf {\bibinfo {volume} {5}},\ \bibinfo {pages} {10742}
  (\bibinfo {year} {2015})}\BibitemShut {NoStop}%
\bibitem [{\citenamefont {Wittmann}\ \emph {et~al.}(2017)\citenamefont
  {Wittmann}, \citenamefont {Maggi}, \citenamefont {Sharma}, \citenamefont
  {Scacchi}, \citenamefont {Brader},\ and\ \citenamefont
  {Marconi}}]{wittmann2017effective}%
  \BibitemOpen
  \bibfield  {author} {\bibinfo {author} {\bibfnamefont {R.}~\bibnamefont
  {Wittmann}}, \bibinfo {author} {\bibfnamefont {C.}~\bibnamefont {Maggi}},
  \bibinfo {author} {\bibfnamefont {A.}~\bibnamefont {Sharma}}, \bibinfo
  {author} {\bibfnamefont {A.}~\bibnamefont {Scacchi}}, \bibinfo {author}
  {\bibfnamefont {J.~M.}\ \bibnamefont {Brader}}, \ and\ \bibinfo {author}
  {\bibfnamefont {U.~M.~B.}\ \bibnamefont {Marconi}},\ }\href@noop {}
  {\bibfield  {journal} {\bibinfo  {journal} {Journal of Statistical Mechanics:
  Theory and Experiment}\ }\textbf {\bibinfo {volume} {2017}},\ \bibinfo
  {pages} {113207} (\bibinfo {year} {2017})}\BibitemShut {NoStop}%
\bibitem [{\citenamefont {Marconi}\ and\ \citenamefont
  {Maggi}(2015)}]{marconi2015towards}%
  \BibitemOpen
  \bibfield  {author} {\bibinfo {author} {\bibfnamefont {U.~M.~B.}\
  \bibnamefont {Marconi}}\ and\ \bibinfo {author} {\bibfnamefont
  {C.}~\bibnamefont {Maggi}},\ }\href@noop {} {\bibfield  {journal} {\bibinfo
  {journal} {Soft Matter}\ }\textbf {\bibinfo {volume} {11}},\ \bibinfo {pages}
  {8768} (\bibinfo {year} {2015})}\BibitemShut {NoStop}%
\bibitem [{\citenamefont {Marconi}\ \emph {et~al.}(2016)\citenamefont
  {Marconi}, \citenamefont {Gnan}, \citenamefont {Paoluzzi}, \citenamefont
  {Maggi},\ and\ \citenamefont {Di~Leonardo}}]{marconi2016velocity}%
  \BibitemOpen
  \bibfield  {author} {\bibinfo {author} {\bibfnamefont {U.~M.~B.}\
  \bibnamefont {Marconi}}, \bibinfo {author} {\bibfnamefont {N.}~\bibnamefont
  {Gnan}}, \bibinfo {author} {\bibfnamefont {M.}~\bibnamefont {Paoluzzi}},
  \bibinfo {author} {\bibfnamefont {C.}~\bibnamefont {Maggi}}, \ and\ \bibinfo
  {author} {\bibfnamefont {R.}~\bibnamefont {Di~Leonardo}},\ }\href@noop {}
  {\bibfield  {journal} {\bibinfo  {journal} {Scientific Reports}\ }\textbf
  {\bibinfo {volume} {6}},\ \bibinfo {pages} {1} (\bibinfo {year}
  {2016})}\BibitemShut {NoStop}%
\end{thebibliography}%





\end{document}